\documentclass[11pt]{article}

\usepackage{amsmath,amssymb,epsfig,color,verbatim,array,bm,bbm,amsfonts,enumerate,enumitem,mathabx, mathtools,lmodern}
\usepackage{graphics,graphicx,float,multirow}
\usepackage[colorlinks=true,citecolor=ForestGreen,urlcolor=black,linkcolor=blue]{hyperref}
\usepackage{blindtext}
\usepackage{microtype}
\usepackage{comment}
\usepackage[utf8]{inputenc}
\usepackage{authblk,booktabs,hhline}
\usepackage[table,dvipsnames]{xcolor}
\usepackage[german,english]{babel}
\usepackage{csquotes}
\usepackage{enumerate}
\usepackage[makeroom]{cancel}
\usepackage{cite}

\definecolor{light-gray}{gray}{0.9}

\usepackage[normalem]{ulem}

\numberwithin{equation}{section}
\allowdisplaybreaks
\let\underbrace\LaTeXunderbrace

\title{Constraint characterization and degree of freedom counting \\
in Lagrangian field theory}

\author[1,2,3]{Ver\'onica Errasti D\'iez \footnote{\href{mailto:vero.erdi@origins-cluster.de}{vero.erdi@origins-cluster.de}}}
\author[4]{Markus Maier \footnote{\href{mailto:markus.maier@hfph.de}{markus.maier@hfph.de}}}
\author[3]{Julio A. M\'endez-Zavaleta \footnote{\href{mailto:julmendez@uv.mx}{julmendez@uv.mx}}}

\affil[1]{ {\small {\it Universit\"{a}ts-Sternwarte,
Fakult\"{a}t f\"{u}r Physik,
Ludwig-Maximilians-Universit\"{a}t M\"{u}nchen,
Scheinerstra\ss e 1,
81679~M\"{u}nchen,
Germany}}}
\affil[2]{ {\small {\it Excellence Cluster ORIGINS,
Boltzmannstra\ss e 2,
85748~Garching bei M\"{u}nchen,
Germany}}}
\affil[3]{ {\small {\it Facultad de F\'isica,
Universidad Veracruzana,
Paseo No. 112,
Desarrollo Habitacional Nuevo Xalapa,
91097~Xalapa-Enr\'iquez,
Mexico}}}
\affil[4]{ {\small {\it Department of Philosophy of Nature and Technology,
Munich School of Philosophy,
Kaulbachstra\ss e 31a,
80539~M\"{u}nchen, Germany}}}

\date{}

\begin{document}

\maketitle

\begin{abstract}
We present a Lagrangian approach to counting
degrees of freedom
in first-order field theories.
The emphasis is on the systematic attainment of
a complete set of constraints.
In particular,
we provide the first comprehensive procedure to ensure
the functional independence of all constraints and
discuss in detail the possible
closures of the constraint algorithm.
We argue degrees of freedom can but need not
correspond to
physical modes.
The appendix comprises
fully worked out, physically relevant examples
of varying complexity.
\end{abstract}

\section{Introduction}
\label{sec:intro}
In most Lagrangian field theories,
there exists a mismatch between
the number of a priori independent field variables 
and the number of degrees of freedom $N_{\textrm{DoF}}$ being propagated.
The determination of $N_{\textrm{DoF}}$ is of crucial importance,
as it directly affects the physics.
More often than not, this
calls for non-trivial analyses.

Within the realm of modified gravity theories, the counting of $N_{\textrm{DoF}}$
has been the subject of vivid interest
in recent years.
Indeed, there exists a plethora of studies in this regard around the Dvali-Gabadadze-Porrati (DGP) model~\cite{Dvali:2000hr},
Galileons~\cite{Nicolis:2008in},
the de~Rham-Gabadadze-Tolley (dRGT) massive gravity~\cite{deRham:2010kj},
bimetric gravity~\cite{Hassan:2011tf,Hassan:2011ea},
beyond Horndeski theories~\cite{Zumalacarregui:2013pma,Gleyzes:2014dya},
Generalized Proca~\cite{Tasinato:2014eka,Heisenberg:2014rta}
and Degenerate-Higher-Order-Scalar-Tensor (DHOST) theories~\cite{BenAchour:2016fzp}, to mention but a few popular settings. The recently proposed
$f(Q)$ gravity~\cite{BeltranJimenez:2017tkd} provides
an example wherein settlement of $N_{\textrm{DoF}}$ remains elusive, see~\cite{DAmbrosio:2023asf} and references therein.

All approaches to the degree of freedom count
developed thus far revolve around constraint algorithms,
whose origin dates back to the work of Dirac  first and Bergmann shortly afterwards. These authors considered a coordinate-dependent approach for autonomous systems within the Hamiltonian formalism. A nice review of the initial proposal
can be found in~\cite{Dirac}.
It was only sensibly later
that the geometrization of the procedure was carried out, yielding the celebrated Presymplectic Constraint Algorithm (PCA)~\cite{GotayNesterHinds1978}.
A different yet equivalent geometric algorithm
was put forward in~\cite{Munoz}.

The PCA was adapted into the Lagrangian formalism stepwise~\cite{GotayNester1979,GotayNester1980,LecandaRomanRoy1992}.
The relation between constraints emerging in the Lagrangian and Hamiltonian algorithms was studied in a coordinate-dependent manner in~\cite{BattlleGomis1986}.
The said relation is based on the so-called
(temporal) evolution operator, whose geometric definition and properties were clarified in~\cite{GraciaPons1989}.

As a remark, we note that even more general geometric constraint algorithms exist.
Prominent examples include
an algorithm for non-autonomous systems in the Lagrangian formalism~\cite{RomanRoy2002}
and algorithms for dynamical systems described implicitly via differential equations~\cite{Gracia1992,Marmo,MunozRomanRoy}.
Unfortunately, there does not seem to exist a complete review of constraint algorithms
and relations between them.

In spite of the vast and sophisticated bodywork on geometric constraint algorithms, such approaches are intrinsically covariant and thus demanding for application in physical theories, where time plays a distinct role. The need to construct multiple auxiliary objects prior to implementation adds to the challenge. This is especially true for elaborate settings, including modified gravity proposals.
The said and kindred hindrances presumably explain (at least in part) the extensive use of the Dirac-Bergmann algorithm, often
supplemented by the Arnowitt-Deser-Misner (ADM) foliation~\cite{Arnowitt:1962hi}, by the gravitational physics community.

In synergy with such fruitful background,
few but noteworthy
Lagrangian algorithms
have been put forward, either
providing or developed as a coordinate-dependent
prescription~\cite{Diaz:2014yua,Diaz:2017tmy,Heidari:2020cil,ErrastiDiez:2020dux}.
Invariably, they aim to facilitate
implementation in exigent theories
and abridge the obtention of physically relevant quantities, such as the number of degrees of freedom $N_{\textrm{DoF}}$.
The present work delves into this line
of research.  

\vspace*{0.5cm}

\noindent
\textbf{Organization}\\
In the main section \ref{sec:method}, we consider first-order Lagrangian field theories and present a coordinate-dependent constraint algorithm for them. Supplemented with information on local (gauge-like) symmetries, the algorithm yields
the significant number $N_{\textrm{DoF}}$ in the theory. Most remarkably,
a thorough procedure for the verification of functional independence among constraints is given
in sections \ref{sec:prim} and \ref{sec:moreit}.
This essential feature had only received modest attention thus far.
Distinct closures of the algorithm are expounded
in section \ref{sec:close}.
For clarity, appendix \ref{app:math} consists of minute implementations of the method in diverse physical theories.
In section \ref{sec:diss}, we reflect on the relation between $N_{\textrm{DoF}}$ and physics.
We draw our conclusions in the final section \ref{sec:concl}.

\vspace*{0.5cm}

\noindent
\textbf{Conventions}\\
We work on a ($d\geq 2$)-dimensional Minkowski manifold $\mathcal{M}$. Spacetime indices are denoted
by the Greek letters $(\mu,\nu,\ldots)$
and raised/lowered with the metric $\eta_{\mu\nu}=\textrm{diag}(-1,1,1,\ldots,1)$
and its inverse $\eta^{\mu\nu}$.
Spacetime coordinates are indicated by
$x^\mu=(x^0,x^1,\ldots,x^{d-1})\equiv (x^0,x^i)$,
with Latin indices $(i,j,\ldots)$ labeling space-like directions.
Dot stands for derivation with respect to time $x^0$.
We employ the short-hands $(\partial_\mu,\partial_i)$ for derivation
with respect to $(x^\mu,x^i)$, respectively.
Summation over repeated indices applies throughout.


\section{Method}
\label{sec:method}

Let $Q^A=Q^A(x^\mu)$
be a finite set of real field variables.
$A=1,2,\ldots, N$ is a collective index,
running over all a priori independent components
of possibly multiple fields of different types.
Consider a first-order classical field theory,
defined by the action
\begin{align}
\label{eq:genact}
S=
\int_{\mathcal{M}}d^d x \, \mathcal{L},
\qquad
\mathcal{L}=
\mathcal{L}(Q^A,\partial_{\mu}Q^A).
\end{align}
Notice we do not consider Lagrangians
with explicit spacetime dependence.
The action (\ref{eq:genact}) may but need not be invariant under local field transformations
of the form
\begin{align}
\label{eq:fieldtrans}
Q^A\rightarrow
Q^A+ \delta_\theta Q^A,
\qquad
\delta_\theta Q^A =
\sum_{p=0}^{P}(-1)^p\left(
\partial_{\mu_1}\partial_{\mu_2}\ldots\partial_{\mu_p}
\theta^M
\right)
({\mathcal{R}_M}^A)^{\mu_1\mu_2\ldots\mu_p},
\end{align}
with $P\in\mathbb{N}_0$.
Here, $\theta^M$ denotes
arbitrary functions of spacetime
labeled by a collective index $M$,
while ${\mathcal{R}_M}^A$ refers to
fixed functions of $(Q^A,\partial_\mu Q^A)$.
All $(\theta^M,{\mathcal{R}_M}^A)$ are taken to be smooth.

Gauge, Lorentz and diffeomorphism transformations
comprise the physically most relevant examples
of (\ref{eq:fieldtrans}).
Discrete, global and conformal (including Weyl)
symmetries of the action
do not affect the degree of freedom count
and hence are not discussed.
Typically,
Lagrangians are postulated 
on the basis of a specific field content $Q^A$
entertaining certain symmetries, if any.
Accordingly and when pertinent,
we assume a priori knowledge of (\ref{eq:fieldtrans}).
Given a Lagrangian
whose symmetries of the relevant form (\ref{eq:fieldtrans})
are unknown a priori,
there exist systematic procedures
to their disclosure~\cite{Banerjee:1999sz,Deriglazov:2009wy,Diaz:2017tmy}.
See also the earlier studies~\cite{Gogilidze:1992ib,Gogilidze:1992jk}.

The number of degrees of freedom
propagated in the theory
can be calculated as~\cite{Diaz:2014yua,Diaz:2017tmy}
\begin{align}
\label{eq:masterfor}
N_{\textrm{DoF}}=N-\frac{1}{2}(g+e+l),
\end{align}
with $l,g,e\in\mathbb{N}_0$.
$g$ is the number of distinct $\theta^M$ functions
in (\ref{eq:fieldtrans}),
while $e\geq g$ is the number of distinct $\theta^M$ functions
plus their successive time derivatives
$(\dot{\theta}^M,\ddot{\theta}^M,\ldots)$
in (\ref{eq:fieldtrans}).
The focus henceforth is on
the number $l$ of
functionally independent
Lagrangian constraints,
iteratively determined as
\begin{align}
\label{eq:lcount}
l=l_1+l_2+\ldots,
\end{align}
where $l_1\geq l_2\geq \ldots$ count primary, secondary, etc.
functionally independent Lagrangian
constraints.

We stress that (\ref{eq:masterfor})
counts degrees of freedom exclusively in terms of Lagrangian parameters. In particular, it
does not require the classification
of constraints into first and second class.
Nonetheless,
such information is not lost, as (\ref{eq:masterfor}) follows
from the map between Hamiltonian and Lagragian parameters~\cite{Diaz:2014yua, Pons:1986zg, Banerjee:1999hu}
\begin{align}
N_1=e, \qquad
N_2=l+g-e,
\end{align}
where $(N_1,N_2)$ denote the number of first and second class constraints in Dirac's canonical formalism, respectively.
Accordingly, (\ref{eq:masterfor})
applies to both point particle systems and classical field theories, with the latter counting degrees of freedom per point in spacetime.
An intrinsically Lagrangian count of degrees of freedom is an important result that was obtained geometrically in~\cite{Diaz:2017tmy} --- see equations (2), (3) and (16) therein.
We remark that (\ref{eq:masterfor}) was derived for first-order theories,
but has also been employed
in higher-order settings~\cite{Klein:2016aiq,Crisostomi:2017aim}.

\subsection{Initialization}
\label{sec:zero}

Consider the (primary) equations of motion of the theory,
in the form
\begin{align}
\label{eq:ELs}
E_A :=
\partial_{\mu}
\left(\frac{\partial\mathcal{L}}{\partial(\partial_{\mu}Q^A)}\right)
-\frac{\partial\mathcal{L}}{\partial Q^A} =
W_{AB}\,\ddot{Q}^B
+U_A=0.
\end{align}  
The (primary) Hessian
\begin{align}
\label{eq:primHessexpr}
W_{AB}=
\partial_{\dot{A}}\partial_{\dot{B}}\mathcal{L}
\end{align}
captures the linear dependence on the generalized accelerations $\ddot{Q}^A$, while 
\begin{align}
\label{eq:Uexpr}
U_A=
\big(\partial_{\dot{A}}\partial_B^{\,i}\mathcal{L}
+\partial_{\dot{B}}\partial_A^{\,i}\mathcal{L}\big)\,
\partial_i \dot{Q}^B
+\big(\partial_A^{\,i}\partial_B^{\,j}\mathcal{L}\big)\,
\partial_i\partial_jQ^B
+\big(\partial_A^{\,\mu}\partial_B\mathcal{L}\big)\,
\partial_\mu Q^B
-\partial_A\mathcal{L}.
\end{align} 
In the above, we have introduced the short-hands
\begin{align}
\label{eq:shortders}
\partial_{\dot{A}}=
\frac{\partial}{\partial \dot{Q}^A},
\qquad
\partial^{\,i}_A=
\frac{\partial}{\partial(\partial_i Q^A)},
\qquad
\partial^{\,\mu}_A=
\frac{\partial}{\partial(\partial_\mu Q^A)},
\qquad 
\partial_A=
\frac{\partial}{\partial Q^A}.
\end{align}

We further rewrite the (primary) equations of motion as
\begin{align}
\label{eq:ELmatrix}
\mathbf{E}^{(1)}:=
\mathbf{W}^{(1)}\ddot{\mathbf{Q}}
+\mathbf{U}^{(1)}=0,
\end{align}
where $(\mathbf{E}^{(1)},\,\ddot{\mathbf{Q}},\,\mathbf{U}^{(1)})$ are $N$-dimensional column vectors and $\mathbf{W}^{(1)}$ is an $N\times N$ square matrix.

\subsection{First iteration}
\label{sec:prim}

The (primary) equations of motion (\ref{eq:ELmatrix})
may but need not encode
second order differential equations (SODEs) in time for all $Q^A$'s.
The number of functionally independent such SODEs
is given by
the (row) rank of the (primary) Hessian $\mathbf{W}^{(1)}$.
The theory may have up to $\big(N-\textrm{rank}\mathbf{W}^{(1)}\big)$ 
primary Lagrangian constraints.

\vspace*{0.5cm}

\noindent
\textbf{\textit{Step I. Rank of the Hessian}} \vspace*{0.2cm}\\
In full generality,
the determination of the (row) rank
of the (primary) Hessian $\mathbf{W}^{(1)}$
is a challenging task.
A conceptually neat and algebraically convenient
manner to do so is as follows.
Assume $\mathbf{W}^{(1)}$ admits left null vectors
\begin{align}
\label{eq:nullcond}
\mathbf{V}^{(1)}\cdot \mathbf{W}^{(1)}=0. 
\end{align}

If no non-trivial solution
to (\ref{eq:nullcond}) exists,
then the (row) rank of the (primary) Hessian
is $N$ and
the theory is said to be regular.
In this case,
the theory possesses
no (primary) Lagrangian constraints $l=l_1=0$.
The constraint determining algorithm
thus terminates.

Else, let $\mathbf{V}^{(1)}$ itself
denote
a maximal set of $M_1\in[1,N)$
linearly independent solutions to (\ref{eq:nullcond}),
normalized as per convenience.
In this case,
the (row) rank of the (primary) Hessian is
$N-M_1$,
the theory is said to be singular
and
up to $M_1$ primary Lagrangian constraints
may exist,
which we proceed to unveil.

\vspace*{0.5cm}

\noindent
\textbf{\textit{Step II. Lagrangian constraints}} \vspace*{0.2cm}\\
Consider the left contraction of
the $M_1$ null vectors $\mathbf{V}^{(1)}$
with the (primary) equations of motion (\ref{eq:ELmatrix})
\begin{align}
\label{eq:primconsgen}
\phi^{(1)}:=
\mathbf{V}^{(1)}\cdot \mathbf{E}^{(1)}=
\mathbf{V}^{(1)}\cdot \mathbf{U}^{(1)}=0.
\end{align}
By definition, $\phi^{(1)}$ is a set of $M_1$ relations
that do not depend on the generalized accelerations $\ddot{Q}^A$.
Any maximal (sub)set
of functionally independent relations
among (\ref{eq:primconsgen})
can be regarded as
the primary Lagrangian constraints
in the theory.
Hence, $l_1\in[0,M_1]$.

It may happen that
all relations in (\ref{eq:primconsgen})
identically vanish.
If so,
there exist no (primary) Lagrangian constraints $l=l_1=0$ and
the constraint algorithm thus terminates. 

In the absence of such trivialization,
the distillation of a maximal (sub)set
of functionally independent relations
from (\ref{eq:primconsgen}) is generally demanding.
A systematic way around the hurdle is similar to
the determination of
the (row) rank of the (primary) Hessian before.
Assume the set of relations in (\ref{eq:primconsgen}),
written as an $M_1$-dimensional column vector $\boldsymbol{\phi}^{(1)}$,
admits solutions to
\begin{align}
\label{eq:nullfunct}
\boldsymbol{\Gamma}^{(1)}\cdot \boldsymbol{\phi}^{(1)}=0.
\end{align}
A generic ansatz for such a vector is
\begin{align}
\label{eq:anothernull}
\boldsymbol{\Gamma}^{(1)}
=(\Gamma^1,\Gamma^2,\ldots,\Gamma^{M_1}),
\qquad 
\Gamma^I=
\Gamma_0^I
+(\Gamma_1^I)^i\,
\partial_i
+(\Gamma_2^I)^{ij}\,
\partial_i\partial_j.
\end{align}
Notice that every component of $\mathbf{\Gamma}^{(1)}$
includes spatial derivative operators up to second order.
In this manner, 
both algebraic and spatial derivative dependences among
the relations in (\ref{eq:primconsgen})
can be identified.

If no non-trivial solution to (\ref{eq:nullfunct}) exists,
then all relations in (\ref{eq:primconsgen})
are functionally independent,
implying $l_1=M_1$.
Moreover, the relations (\ref{eq:primconsgen}) themselves
can be interpreted as
the primary Lagrangian constraints
in the theory.

Else, let $\boldsymbol{\Gamma}^{(1)}$ itself
denote a maximal set of $m_1\in[1,M_1)$
linearly independent solutions 
to (\ref{eq:nullfunct}),
normalized as per convenience.
In this case, there exist $l_1=M_1-m_1$
primary Lagrangian constraints in the theory,
which can be parametrized as
\begin{align}
\label{eq:indeplagcons}
\phi_\ast^{(1)}:=\big(\boldsymbol{\Gamma}_0^{(1)}\big)^\perp
\cdot \boldsymbol{\phi}^{(1)}=0,
\end{align}
where $\big(\boldsymbol{\Gamma}_0^{(1)}\big)^\perp$
stands for a maximal set of linearly independent row vectors orthogonal to
\begin{align}
\label{eq:Gamma0}
\boldsymbol{\Gamma}_0^{(1)}=(
\Gamma^1_0,\Gamma^2_0,\ldots,\Gamma^{M_1}_0)
\subseteq \boldsymbol{\Gamma}^{(1)},
\end{align}
normalized as per convenience\footnote{Toy models
for functional dependence detection
and characterization by the above procedure
are given in appendix \ref{sec:funcindep}.\label{footnote1}}.

To sum up, for $l_1\neq0$,
let 
\begin{align}
\label{eq:primcons}
\Phi^{(1)}=
\left\{
\begin{array}{llll}
\phi^{(1)} 
\textrm{ \hspace*{0.1cm} in (\ref{eq:primconsgen})}
&
\textrm{ if } l_1=M_1,
\vspace*{0.2cm}\\
\phi_\ast^{(1)}
\textrm{ \hspace*{0.1cm} in (\ref{eq:indeplagcons})} 
&
\textrm{ if } l_1=M_1-m_1
\end{array}
\right.
\end{align}
parametrize the
primary Lagrangian constraints
in the theory.
For later convenience,
let $\boldsymbol{\Phi}^{(1)}$
denote those same primary Lagrangian constraints,
written as an $l_1$-dimensional column vector.

\vspace*{0.5cm}

\noindent
\textbf{\textit{Step III. Stability of
the Lagrangian constraints}}
\vspace*{0.2cm}\\
Let $R=1,2,\ldots,l_1$
label the primary Lagrangian constraints
(\ref{eq:primcons}).
Self-consistency of the theory
under time evolution
implies 
\begin{align}
\label{eq:primtime}
E_R\equiv\dot{\Phi}_R^{(1)}:=
W_{RA}\ddot{Q}^A
+U_R
=0
\end{align}
must hold true,
where 
\begin{align}
\label{eq:hessUtilde}
W_{RA}=
\partial_{\dot{A}}\Phi^{(1)}_R
+\big(\partial_{\dot{A}}^{\, i}\Phi^{(1)}_R\big)\,
\partial_i,
\qquad
U_R=
\big(\partial_A^{\, ij}\Phi^{(1)}_R\big)\,
\partial_i\partial_j\dot{Q}^A
+\big(\partial_A^{\, i}\Phi^{(1)}_R\big)\,
\partial_i\dot{Q}^A
+\big(\partial_A\Phi^{(1)}_R\big)\,
\dot{Q}^A.
\end{align}
It is convenient to rewrite the above as
\begin{align}
\label{eq:EL2matrix}
\mathbf{E}^{(2)}\equiv
\dot{\boldsymbol{\Phi}}^{(1)}:=
\mathbf{W}^{(2)}\ddot{\mathbf{Q}}
+\mathbf{U}^{(2)}=0,
\end{align}
where $(\mathbf{E}^{(2)},\,\mathbf{U}^{(2)})$
are $l_1$-dimensional column vectors and $\mathbf{W}^{(2)}$ is an $l_1\times N$ rectangular matrix.

The $l_1$ conditions (\ref{eq:primtime}),
viewed as
the secondary equations of motion
for the system (\ref{eq:EL2matrix}),
may give rise to up to $l_1$
secondary Lagrangian constraints,
which are to be unveiled in
a second iteration
of the constraint algorithm.
Indeed,
a conceptually simple
repetition of
the just described procedure
yields the secondary Lagrangian constraints
in the theory, if any.
The only formal subtlety amounts to
ensuring that only functionally independent
secondary Lagrangian constraints
are considered
in the degree of freedom count
(\ref{eq:masterfor}).
To this aim, let
\begin{align}
\label{eq:secELs}
\mathbf{E}^{(2)\downarrow}:=
\mathbf{W}^{(2)\downarrow}\ddot{\mathbf{Q}}
+\mathbf{U}^{(2)\downarrow}
=0
\end{align}
encompass
the primary (\ref{eq:ELmatrix})
and secondary (\ref{eq:EL2matrix})
equations of motion, in the form
\begin{align}
\label{eq:secHessU}
\mathbf{E}^{(2)\downarrow}=
\left(
\begin{array}{ccc}
\mathbf{E}^{(1)} \\
\mathbf{E}^{(2)}
\end{array}
\right),
\qquad
\mathbf{W}^{(2)\downarrow}=
\left(
\begin{array}{ccc}
\mathbf{W}^{(1)} \\
\mathbf{W}^{(2)}
\end{array}
\right),
\qquad
\mathbf{U}^{(2)\downarrow}=
\left(
\begin{array}{ccc}
\mathbf{U}^{(1)} \\
\mathbf{U}^{(2)}
\end{array}
\right).
\end{align}
Here,
$\mathbf{E}^{(2)\downarrow}$ and 
$\mathbf{U}^{(2)\downarrow}$ are
$(N+l_1)$-dimensional column vectors, 
while $\mathbf{W}^{(2)\downarrow}$ is
an $(N+l_1)\times N$ rectangular matrix.
The second iteration
in the constraint algorithm
takes $\mathbf{E}^{(2)\downarrow}$
as a starting point,
as opposed to merely $\mathbf{E}^{(2)}$.

\subsection{Generic iteration $n\geq2$}
\label{sec:moreit}
For $m<n$,
let $\boldsymbol{\Phi}^{(m)}$,
denote the $m$-stage Lagrangian constraints,
written as an $l_m$-dimensional column vector.
Then, let
\begin{align}
\label{eq:consstack}
\boldsymbol{\Phi}^{(n-1)\downarrow}:=
\left(
\begin{array}{ccc}
\boldsymbol{\Phi}^{(1)} \\
\boldsymbol{\Phi}^{(2)} \\
\vdots \\
\boldsymbol{\Phi}^{(n-1)}
\end{array}
\right)
\end{align}
denote the ordered collection of all
Lagrangian constraints unveiled thus far.
We stress that 
$\boldsymbol{\Phi}^{(n-1)\downarrow}$ is
an $(N^+-N)$-dimensional column vector
whose components
have already been proven
functionally independent,
where
\begin{align}
\label{eq:defNpm}
N^+=N+\sum_{p=1}^{n-1}l_p.
\end{align}

On the other hand, let
\begin{align}
\label{eq:nmatrixELs}
\mathbf{E}^{(n)}\equiv
\dot{\boldsymbol{\Phi}}^{(n-1)}:=
\mathbf{W}^{(n)}\cdot \ddot{\mathbf{Q}}
+\mathbf{U}^{(n)}=0
\end{align}
denote the $n$-stage equations of motion,
written as an $(l_{n-1})$-dimensional column vector.
Notice that  
the $n$-stage Hessian $\mathbf{W}^{(n)}$ is
an $(l_{n-1}\times N)$ rectangular matrix.
Further, let
\begin{align}
\label{eq:ndownmatrixELs}
\mathbf{E}^{(n)\downarrow}:=
\mathbf{W}^{(n)\downarrow}\cdot \ddot{\mathbf{Q}}
+\mathbf{U}^{(n)\downarrow}=0
\end{align}
denote the ordered collection of
all equations of motion
up to and including the $n$-stage,
in the form
\begin{align}
\label{eq:ELnarrow}
\mathbf{E}^{(n)\downarrow}=
\left(
\begin{array}{ccc}
\mathbf{E}^{(1)} \\
\mathbf{E}^{(2)} \\
\vdots \\
\mathbf{E}^{(n)}
\end{array}
\right),
\qquad
\mathbf{W}^{(n)\downarrow}=
\left(
\begin{array}{ccc}
\mathbf{W}^{(1)} \\
\mathbf{W}^{(2)} \\
\vdots \\
\mathbf{W}^{(n)}
\end{array}
\right),
\qquad
\mathbf{U}^{(n)\downarrow}=
\left(
\begin{array}{ccc}
\mathbf{U}^{(1)} \\
\mathbf{U}^{(2)} \\
\vdots \\
\mathbf{U}^{(n)}
\end{array}
\right).
\end{align}
Here, $(\mathbf{E}^{(n)\downarrow},\,\mathbf{U}^{(n)\downarrow})$
are $N^+$-dimensional column vectors,
while $\mathbf{W}^{(n)\downarrow}$
is an $N^+\times N$ rectangular matrix. 

\vspace*{0.5cm}

\noindent
\textbf{\textit{Step I. Rank of the Hessian}} \vspace*{0.2cm}\\
First, the row rank of
$\mathbf{W}^{(n)\downarrow}$
is to be determined.
To this aim,
assume it admits left null vectors
\begin{align}
\label{eq:nnullhess}
\mathbf{V}^{(n)\downarrow}\cdot\mathbf{W}^{(n)\downarrow}=0.
\end{align}
A generic ansatz for such a null vector is\footnote{Examples requiring the postulation and calculation of a nonobvious vector $\mathbf{V}^{(2)\downarrow}$ are given in appendices \ref{sec:podol} and \ref{sec:MMEPN}.}
\begin{align}
\label{eq:gennullnit}
\mathbf{V}^{(n)\downarrow}\equiv 
(\mathbf{V}^{(1)},
\mathbf{V}^{(2)},
\ldots,
\mathbf{V}^{(n)}),
\qquad
\mathbf{V}^{(m)}=
\mathbf{V}_0 +
\sum_{p=1}^{n-m}
(\mathbf{V}_p)^{i_1\ldots i_p} \partial_{i_1}\ldots \partial_{i_p}.
\end{align}
By construction,
there indeed exist solutions to (\ref{eq:nnullhess}):
they trivially extend
the left null vector(s) $\mathbf{V}^{(n-1)\downarrow}$ found in the
immediately previous
stage
through $\mathbf{V}^{(n)}=0$.
Such solutions
do not carry new information.
Consequently, they are to be dismissed.

If no left null vector to
$\mathbf{W}^{(n)\downarrow}$
exists
such that $\mathbf{V}^{(n)}\neq0$,
then 
\begin{align}
\label{eq:rankcoll}
\textrm{rrank}\left(\mathbf{W}^{(n)\downarrow}\right)
=\textrm{rrank}\left(\mathbf{W}^{(n-1)\downarrow}\right)
+l_{n-1},
\end{align}
where rrank stands for row rank.
It follows that
$l_n=0$ and the algorithm thus terminates.
In this case, $l=N^+-N$.

Else,
let $\mathbf{V}^{(n)\downarrow}$ itself
denote a maximal set of $M_n\in[1,l_{n-1}]$
linearly independent left null vectors
to $\mathbf{W}^{(n)\downarrow}$
such that $\mathbf{V}^{(n)}\neq0$,
normalized as per convenience.
In this case,
\begin{align}
\label{eq:ranknocoll}
\textrm{rrank}\left(\mathbf{W}^{(n)\downarrow}\right)
=\textrm{rrank}\left(\mathbf{W}^{(n-1)\downarrow}\right)
+l_{n-1}-M_n,
\end{align}
implying that up to $M_n$
$n$-stage Lagrangian constraints may exist.

Algebraic ease dictates that the disclosure
of the $n$-stage Lagrangian constraints,
if any,
is carried out in two steps.
The first step
guarantees functional independence
within the $n$th iteration.
The second step
guarantees functional independence
with respect to previous iterations.

\vspace*{0.5cm}

\noindent
\textbf{\textit{Step II. Lagrangian constraints}}\\
\textit{Substep IIA. Functional independence within the stage}
\vspace*{0.2cm} \\
Consider the set of $M_n$ relations
\begin{align}
\label{eq:rawconsgen}
\phi^{(n)}:=
\mathbf{V}^{(n)\downarrow}\cdot
\mathbf{E}^{(n)\downarrow}=
\mathbf{V}^{(n)\downarrow}\cdot
\mathbf{U}^{(n)\downarrow}=0.
\end{align}
If the above relations trivially vanish,
there exist no $n$-stage Lagrangian constraints
$l_n=0$.
The constraint algorithm thus terminates,
yielding $l=N^+-N$.

Else,
a maximal (sub)set of
functionally independent relations
among (\ref{eq:rawconsgen})
is to be extracted.
To this aim,
let $\boldsymbol{\phi}^{(n)}$
denote the relations (\ref{eq:rawconsgen}),
written as an $M_n$-dimensional column vector.
Assume $\boldsymbol{\phi}^{(n)}$ admits solutions to
\begin{align}
\label{eq:nPhisnull}
\mathbf{\Gamma}^{(n)}\cdot 
\boldsymbol{\phi}^{(n)}=0.
\end{align}
A generic ansatz for such a vector is
\begin{align}
\label{eq:Gammancomp}
\mathbf{\Gamma}^{(n)}=
(\Gamma^1,\Gamma^2,\ldots,\Gamma^{M_n}),
\qquad
\Gamma^I=\Gamma_0^I+\sum_{p=1}^{n-1}(\Gamma_p^I)^{i_1\ldots i_p}\,
\partial_{i_1}\ldots\partial_{i_p}.
\end{align}

If no non-trivial
solution to (\ref{eq:nPhisnull})
exists, then
all relations in (\ref{eq:rawconsgen})
are functionally independent among themselves.

Else, let
$\mathbf{\Gamma}^{(n)}$ itself
denote
a maximal set of $m_n\in[1,M_n)$
linearly independent solutions to (\ref{eq:nPhisnull}),
normalized as per convenience.
In this case,
a maximal subset of $M_n-m_n$ functionally independent relations among (\ref{eq:rawconsgen})
is 
\begin{align}
\label{eq:indepncons}
\phi^{(n)}_\ast:=
\big(\mathbf{\Gamma}^{(n)}_0\big)^\perp
\cdot \boldsymbol{\phi}^{(n)}=0,
\end{align}
where $\big(\mathbf{\Gamma}^{(n)}_0\big)^\perp$ stands for a maximal set of 
linearly independent row vectors orthogonal to
\begin{align}
\label{eq:perpnGamma}
\boldsymbol{\Gamma}_0^{(n)}=(
\Gamma^1_0,\Gamma^2_0,\ldots,\Gamma^{M_n}_0)
\subseteq \boldsymbol{\Gamma}^{(n)},
\end{align}
normalized as per convenience.\textsuperscript{\ref{footnote1}}

In conclusion,
let
\begin{align}
\label{eq:nrelsindep}
\boldsymbol{\varphi}^{(n)}=
\left\{
\begin{array}{llll}
\boldsymbol{\phi}^{(n)}
\textrm{ \hspace*{0.1cm} in (\ref{eq:rawconsgen})}
&
\textrm{ if (\ref{eq:nPhisnull})
does not admit non-trivial solutions,} \\
\boldsymbol{\phi}^{(n)}_\ast
\textrm{ \hspace*{0.1cm} in (\ref{eq:indepncons})} 
&
\textrm{ otherwise }
\end{array}
\right.
\end{align}
denote a maximal (sub)set of functionally independent relations among (\ref{eq:rawconsgen}).
Let $\mathfrak{M}_n$ denote their number,
where $\mathfrak{M}_n=M_n$ or $M_n-m_n$,
as per (\ref{eq:nrelsindep}).
For subsequent convenience, let $\boldsymbol{\varphi}^{(n)}$
denote those same relations,
arranged in a column vector
of dimension $\mathfrak{M}_n$.

\vspace*{0.5cm}

\noindent
\textit{Substep IIB. Functional independence
with respect to previous stages}
\vspace*{0.2cm}\\
In an $n\geq 2$ iteration of the constraint algorithm,
the disclosed maximal (sub)set of
functionally independent relations
(\ref{eq:nrelsindep})
cannot be immediately
regarded as parametrizing the
$n$-stage Lagrangian constraints.
This is because
(\ref{eq:nrelsindep})
is not necessarily
functionally independent from
the Lagrangian constraints 
unveiled in previous iterations.
We proceed to ensure such
retroactive functional independence.

Let
\begin{align}
\label{eq:auxncons}
\boldsymbol{\Psi}^{(n)\downarrow}:=
\left(
\begin{array}{ccc}
\boldsymbol{\Phi}^{(n-1)\downarrow} \\
\boldsymbol{\varphi}^{(n)}
\end{array}
\right)
\end{align}
denote the ordered collection of
all previous stages' Lagrangian constraints $\boldsymbol{\Phi}^{(n-1)\downarrow}$ in
(\ref{eq:consstack}) and 
the relations $\boldsymbol{\varphi}^{(n)}$ in (\ref{eq:nrelsindep}).
Recall that, by construction,
both distinct sets 
$\boldsymbol{\Phi}^{(n-1)\downarrow}$
and $\boldsymbol{\varphi}^{(n)}$
comprise only functionally independent relations.
As a result, upon joint consideration,
\begin{align}
\label{eq:rrankPsivalues}
l_n=
\textrm{rrank}\left(
\boldsymbol{\Psi}^{(n)\downarrow}
\right) - \textrm{rrank}\left(
\boldsymbol{\Phi}^{(n-1)\downarrow}
\right) \in [0,\mathfrak{M}_n].
\end{align}

In order to determine $l_n$, assume $\boldsymbol{\Psi}^{(n)\downarrow}$
admits solutions to
\begin{align}
\label{eq:nullvarphi}
\boldsymbol{\Upsilon}^{(n)\downarrow}\cdot
\boldsymbol{\Psi}^{(n)\downarrow}=0.
\end{align}
A generic ansatz for such a vector is\footnote{An example requiring the postulation and calculation of a nonobvious vector $\boldsymbol{\Upsilon}^{(2)\downarrow}$ is given in appendix \ref{sec:palatini}.}
\begin{align}
\label{eq:gammaans}
\boldsymbol{\Upsilon}^{(n)\downarrow}=
(\boldsymbol{\Upsilon}^{(1)},
\boldsymbol{\Upsilon}^{(2)},\ldots,
\boldsymbol{\Upsilon}^{(n)}),
\qquad 
\boldsymbol{\Upsilon}^{(m)}=
\boldsymbol{\Upsilon}_0^{(m)}+
\sum_{p=1}^{n-m}
\big(\boldsymbol{\Upsilon}_p^{(m)}\big)^{i_1\ldots i_p} \partial_{i_1}\ldots \partial_{i_p}.
\end{align}

If a maximal set of $\mathfrak{M}_n$ 
linearly independent solutions to
(\ref{eq:nullvarphi}) exists, then
all relations (\ref{eq:nrelsindep})
are functionally dependent
with respect to $\boldsymbol{\Phi}^{(n-1)\downarrow}$.
In this case,
(\ref{eq:auxncons}) has the minimal row rank
\begin{align}
\label{eq:rrankPsi}
\textrm{rrank}\left(
\boldsymbol{\Psi}^{(n)\downarrow}
\right)=\textrm{rrank}\left(
\boldsymbol{\Phi}^{(n-1)\downarrow}
\right)=N^+-N
\end{align}
and therefore
$l_n=0$.
The constraint algorithm thus terminates,
yielding $l=N^+-N$.

Else, start by considering the diametrically opposite instance.
If no non-trivial solution to
(\ref{eq:nullvarphi}) exists,
then the relations (\ref{eq:nrelsindep})
are functionally independent
with respect to $\boldsymbol{\Phi}^{(n-1)\downarrow}$.
In this case, $l_n=\mathfrak{M}_n$.
Moreover, the relations
(\ref{eq:nrelsindep}) can
be regarded as the $n$-stage Lagrangian constraints in the theory.

Next, consider all intermediate instances.
Let $\boldsymbol{\Upsilon}^{(n)\downarrow}$ itself
denote a maximal set of $\mathfrak{m}_n\in[1,\mathfrak{M}_n)$
linearly independent solutions to
(\ref{eq:nullvarphi}),
normalized as per convenience.
In this case,
$l_n=\mathfrak{M}_n-\mathfrak{m}_n$ and
a maximal subset of  functionally independent relations among (\ref{eq:nrelsindep})
is 
\begin{align}
\label{eq:indepncons_n}
\varphi^{(n)}_\ast:=
\big(\boldsymbol{\Upsilon}^{(n)\downarrow}_0\big)^\perp
\cdot \boldsymbol{\Psi}^{(n)\downarrow}=0,
\end{align}
where $\big(\boldsymbol{\Upsilon}^{(n)\downarrow}_0\big)^\perp$ stands for a maximal set of  linearly independent row vectors orthogonal to
\begin{align}
\label{eq:Upsilon0}
\boldsymbol{\Upsilon}_0^{(n)\downarrow}=(
\boldsymbol{\Upsilon}_0^{(1)},\boldsymbol{\Upsilon}_0^{(2)},\ldots,\boldsymbol{\Upsilon}_0^{(n)})
\subseteq \boldsymbol{\Upsilon}^{(n)\downarrow},
\end{align}
normalized as per convenience.

In short, for $l_n\neq 0$,
let 
\begin{align}
\label{eq:nstagecons}
\Phi^{(n)}=
\left\{
\begin{array}{llll}
\varphi^{(n)}
\textrm{ \hspace*{0.1cm} in (\ref{eq:nrelsindep})}
&
\textrm{ if (\ref{eq:nullvarphi}) does not admit non-trivial solutions,} \\
\varphi^{(n)}_\ast
\textrm{ \hspace*{0.1cm} in (\ref{eq:indepncons_n})} 
&
\textrm{ otherwise } 
\end{array}
\right.
\end{align}
denote the
$n$-stage Lagrangian constraints.

\vspace*{0.5cm}

\noindent
\textbf{\textit{Step III. Stability of
the Lagrangian constraints.}}
\vspace*{0.2cm}\\
Self-consistency demands that
the
$n$-stage Lagrangian constraints
are preserved under time evolution:
$\dot{\Phi}^{(n)}=0$, for all $l_n$ relations
in (\ref{eq:nstagecons}).
This condition may yield
up to $l_n$
Lagrangian constraints
in a subsequent iteration
of the constraint algorithm
\subsection{Closure}
\label{sec:close}

In order to unequivocally establish
the number of degrees of freedom $N_{\textrm{DoF}}$ in a given theory,
it is imperative to pursue
any constraint algorithm to its closure.
Unfortunately and especially within coordinate-dependent Lagrangian approaches,
persistence to termination is not always the case,
as alerted against in~\cite{ErrastiDiez:2020dux,Janaun:2023nxz}.
In the method just advocated,
there exist 3 distinct manners
in which the constraint algorithm may close.

Let $n_{\textrm{f}}$ denote the (finite) final iteration, wherein $l_{n_{\textrm{f}}}=0$.
As per (\ref{eq:nmatrixELs}), let
$\mathbf{E}^{(n_{\textrm{f}})}$
and $\mathbf{W}^{(n_{\textrm{f}})}$ 
denote
the associated $n_{\textrm{f}}$-stage equations of motions
and Hessian, respectively. 
$l_{n_{\textrm{f}}}=0$ is a direct consequence of one of the following instances.

\begin{enumerate}[label=\textcolor{blue}{\arabic*.}, ref=\arabic*]

\item 
\label{item:Wclose}
$\mathbf{W}^{(n_{\textrm{f}})}$
has maximal row rank (\ref{eq:rankcoll}).
In this case,
consistency under time evolution of
the $(n_{\textrm{f}}-1)$-stage Lagrangian constraints
is ensured dynamically,
through second-order (in time) differential equations of the variables $Q^A$.
Examples of this closure
can be found in appendices
\ref{sec:chiral} and
\ref{sec:proca}--\ref{sec:MMEPN}.

\item
\label{item:trivialclose}
$\mathbf{W}^{(n_{\textrm{f}})}$
does not have maximal row rank, but
all contractions of
its chosen left null vectors
with  $\mathbf{E}^{(n_{\textrm{f}})}$
identically vanish. Namely,
(\ref{eq:rawconsgen}) is identically satisfied.
This closure is the trivial expression
of the functional dependence of 
the would-be Lagrangian constraints arising at the $n_{\textrm{f}}$ stage
on the previous stages' Lagrangian constraints.
An example of this closure is provided in appendix \ref{sec:maxwell}.

\item
\label{item:hardclose} Equation
(\ref{eq:rawconsgen}) is not identically satisfied,
but it exclusively comprises
relations that are functionally dependent
on the previous stages' Lagrangian constraints.
Namely,
(\ref{eq:rrankPsi})  is fulfilled.
This constitutes the non-trivial counterpart to the previously described closure.
An example is given in appendix \ref{sec:palatini}.
\end{enumerate}

We are not aware of any physical example
within the scope of this work
where constraint algorithms fail to close
at a finite number of iterations.

\subsection{Remarks}
\label{sec:rem}

Lagrangian constraints are not uniquely defined,
only the space they span is.
At every iteration,
we have advocated for
the most convenient choice.
Such choice is model-dependent.

For all iterations in
the constraint algorithm,
it has been implicitly assumed that
the row rank of the relevant Hessian
remains constant.
Presumably,
the distinct dynamical behavior
of the field configuration(s) for which
there is a change in the rank
of one or more of the Hessians
remains encoded in the final stage's stack of equations of motion $\mathbf{E}^{(n_{\textrm{f}})\downarrow}$ --- and hence in the later discussed dynamical problem (\ref{eq:indepEls}).

The method readily applies to
higher-order field theories whose
equations of motion are linear in the generalized accelerations $\ddot{Q}^A$, as in (\ref{eq:ELmatrix}).
In this case, the (primary) Hessian
and remaining terms have a more complicated,
order-dependent relation to the Lagrangian than (\ref{eq:primHessexpr}) and (\ref{eq:Uexpr}),
but the algorithm per se remains unaltered.

\section{Dynamical versus physical modes}
\label{sec:diss}

Degrees of freedom are 
a foundational subject in physics.
As such, they drive sustained investigations
on and around themselves.
Prominent questions under survey include their very definition,
the well-posedness and solvability of
their associated dynamical equations and their relation
to physical observables.
In this section,
we briefly ponder on such
open-ended
problems.

It is customary to view $N_{\textrm{DoF}}$
as counting the (pairs of) initial conditions
needed to define the dynamical problem of a given theory. 
We proceed to elucidate the previous assertion within the scope of section \ref{sec:method}. This allows us to confront dynamical and physical modes.

Consider a theory of the form (\ref{eq:genact})
for which the constraint algorithm has been successfully pursued until closure $l_{n_{\textrm{f}}} = 0$.
As per (\ref{eq:ndownmatrixELs}) and (\ref{eq:ELnarrow}), let
$\mathbf{E}^{(n_{\textrm{f}})\downarrow}$
and
$\mathbf{W}^{(n_{\textrm{f}})\downarrow}$
denote
the exhaustive stack of
associated equations of motions
and Hessians, respectively. 

By construction and
for singular theories,
$\mathbf{W}^{(n_{\textrm{f}})\downarrow}$
is an $N^+_{\textrm{f}}\times N$
matrix with non-maximal row rank
\begin{align}
\label{eq:rrankHessnf}
\varrho\equiv
\textrm{rrank}\left(
\mathbf{W}^{(n_{\textrm{f}})\downarrow}
\right)=
N_{\textrm{f}}^+- M_{\textrm{tot}}<N^+_{\textrm{f}},
\qquad N^+_{\textrm{f}}=
N+l,
\qquad
M_{\textrm{tot}}=\sum_{p=1}^{n_{\textrm{f}}} M_p.
\end{align}
(Recall that $M_n$ denotes the maximal number
of linearly independent left null vectors to
$\mathbf{W}^{(n_{\textrm{f}})\downarrow}$ that
necessarily and at most involve the $n$-stage Hessian $\mathbf{W}^{(n)}$.)
It follows that 
$\mathbf{W}^{(n_{\textrm{f}})\downarrow}$
admits a maximal set of $M_{\textrm{tot}}$ linearly independent left null vectors
$\mathbf{V}^{(n_{\textrm{f}})\downarrow}$,
normalized as per convenience.
Let $(\mathbf{V}^{(n_{\textrm{f}})\downarrow})^\perp$
denote a maximal set of $\varrho$ linearly independent
row vectors orthogonal to $\mathbf{V}^{(n_{\textrm{f}})\downarrow}$,
normalized as per convenience.

Consider
\begin{align}
\label{eq:indepEls}
\mathbb{E}:=
(\mathbf{V}^{(n_{\textrm{f}})\downarrow})^\perp\cdot \mathbf{E}^{(n_{\textrm{f}})\downarrow}=0.
\end{align}
The above comprises $\varrho$ functionally independent
second order differential equations (SODEs)
in time for (some of) the variables $Q^A$.
Supplemented by $2N_{\textrm{DoF}}$ initial values for $(Q^A,\dot{Q}^A)$,
(\ref{eq:indepEls})
defines the dynamic problem of the theory
(\ref{eq:genact}).
Conversely, $N_{\textrm{DoF}}$ counts the pairs $(Q^A,\dot{Q}^A)$
whose time dependence is encoded in 
the just described dynamical problem.
In other words, $N_{\textrm{DoF}}$ counts
dynamical modes.
If any, variables $Q^A$ present in the Lagrangian (\ref{eq:genact}) but not determined by the dynamical problem comprise pure gauge modes.
In this regard,
the interested reader is gladly referred
to~\cite{Golovnev:2022rui}.

The dynamical problem may but need not fall within the scope of  
the Cauchy-Kovalevskaya (CK) theorem.
When it does,
a unique analytical solution is guaranteed to exist.
When it does not,
existence (let alone uniqueness)
of analytical solutions
cannot be generically ascertained.
Physics-driven extensions to the CK theorem
thus comprise an enticing line of mathematical research.

Consider a generic Lagrangian field theory,
possibly beyond the subclass in section \ref{sec:method}.
Presume
a fortunate case
in which one or more analytical solutions to its dynamical problem can be found.
Even then, the obtained dynamical modes
should not be immediately identified with physical modes,
in the sense that
dynamical modes may exhibit a behavior that is incompatible with well-established physical principles and/or observation.
Reasons are plentiful.

First, consider stability criteria.
For instance, solutions
could be perturbatively unstable,
in the sense of lacking robustness
against small deviations in the initial data
and/or free parameters.
Whenever
in conflict
with observation,
such solutions are to be
disregarded.
A lucid introduction to
the most frequent perturbative instabilities in gravitational settings is~\cite{Delhom:2022vae}.
Numerical examples of critical values for free parameters which dramatically
destabilize a theory can be found in~\cite{Pavsic:2013noa}.
Moreover, it has been known for a while that
instabilities could also appear only at the non-perturbative level~\cite{Smilga:2004cy}.
For an enlightening recent review apropos, see~\cite{Damour:2021fva}.

Causality is yet another essential
requirement for a dynamical mode
to be deemed physical.
As a remarkable example,
we note~\cite{Hassan:2017ugh},
where causality was employed to constrain dynamical modes for certain massive gravity theories.
Overall, the quest remains for
necessary and sufficient conditions
that guarantee physicality of dynamical modes,
even within the theoretical realm.

More generally,
it is worth considering physicality of a (classical) field theory as a whole.
From a philosophical perspective, 
degrees of freedom might help to address the question of  physical equivalence between theories,
which enjoys a long tradition in the philosophy of science, see e.g.~\cite{vanFraassen.1986,Quine.1975,Qadir.1976}.
While several notions of equivalence are discussed in the recent literature~\cite{Weatherall.2019a,Weatherall.2019b}, there exists widespread consensus that dynamical equivalence of two theories is a necessary condition for their empirical equivalence~\cite{Wolf.2023}.
In this context, we regard a match in the number of degrees of freedom 
as a prefatory necessary (but not sufficient) condition
for physical equivalence
between two theories.

\section{Conclusions}
\label{sec:concl}

We have presented
a Lagrangian method to count degrees of freedom in first-order classical field theories.
The emphasis is two-fold.
First, a systematic and algebraically convenient procedure to establish the functional independence of the constraints that may be present in such  theories.
Second, a detailed discussion on
the possible closures of the associated constraint algorithm.
Both are consequential aspects
that are rarely explicitly addressed in akin
Lagrangian approaches. 
Non-exhaustive counterexamples to the former omission can be found in~\cite{Diaz:2017tmy,ErrastiDiez:2020dux}.
The latter was painstakingly discussed in~\cite{ErrastiDiez:2020dux}
and, for a certain family of massive electrodynamics theories, in~\cite{Janaun:2023nxz}.

We stress that functional independence among constraints is essential to the postulation of self-consistent theories. This is particularly relevant when a given constraint structure is being pursued, i.e.~fixed values for the number of primary, secondary, etc. (Lagrangian) constraints. Against this background, we note that a suitable (row) rank reduction of an $n$-stage Hessian does not ensure the desired number of $n$-stage constraints are generated.
While such (row) rank reduction is a necessary condition
for the sought constraint structure, 
it is premature to regard it as sufficient
on its own~\cite{Janaun:2023nxz}. 

Failure to close a constraint algorithm may yield an incorrect number of degrees of freedom $N_{\textrm{DoF}}$.
It could happen that 
$N_{\textrm{DoF}}$ is overestimated, via the overlooking
of overconstrained systems.
Contrastively, $N_{\textrm{DoF}}$ may be underestimated, via the misidentification of functionally dependent relations of the type in
(\ref{eq:rawconsgen}) as (Lagrangian) constraints.

Last but not least, we have discussed several non-trivial conditions
that propagating degrees of freedom must fulfill before they can be regarded as physical modes.

\vspace*{0.5cm}

\noindent
{\bf Acknowledgments.}
We are greatly indebted to 
Jordi Gaset Rif\`a,
for enlightening discussions
and constructive feedback on
an earlier version of this manuscript,
and to Narciso Rom\'{a}n-Roy,
for his comprehensive and cogent review of constraint algorithms.
VED is grateful to all members of the Faculty of Physics
at Universidad Veracruzana
for their hospitality during the final stages of this work; in particular,
members of \emph{L\'inea de Generaci\'on y Aplicaci\'on del Conocimiento, Geometr\'ia y Gravitaci\'on}. MM would like to thank Benjamin Rathgeber for his constant advice and his sincere encouragement to pursue an interdisciplinary line of research.
The work of VED is funded by
the Deutsche Forschungsgemeinschaft
(DFG, German Research Foundation)
under Germany's Excellence Strategy -- EXC-2094 -- 390783311. JAMZ is funded by CONAHCYT's
\emph{Estancias Posdoctorales por M\'exico} grant no.~898686.

\appendix

\section{Appendix: Examples}
\label{app:math}

We begin in section \ref{sec:funcindep}, by providing
simple yet illuminating examples
for the obtention of functionally independent constraints from a given out-of-context set.
The remainder of the appendix is devoted
to the explicit application of
the method presented in section \ref{sec:method}  to count degrees of freedom in
various examples of physical relevance.
The appendix thus serves to amply illustrate
the use of the method,
at various levels of algebraic intricacy.

\vspace*{0.5cm}

\noindent
\textbf{Notation.}
Brackets denoting
symmetrization and antisymmetrization of indices are defined as
$T_{(\mu\nu)}=(T_{\mu\nu}+T_{\nu\mu})/2$ and 
$T_{[\mu\nu]}=(T_{\mu\nu}-T_{\nu\mu})/2$,
respectively.
For the 2-dimensional examples in sections \ref{sec:funcindep}, \ref{sec:chiral}, and \ref{sec:MMEPN}, 
we use the short hand $T^\prime=\partial_1T$.
Natural units are employed throughout.

\subsection{Detection and avoidance of functional dependence among ad hoc constraints}
\label{sec:funcindep}

\textit{Toy model I}\\
Consider a theory of the form (\ref{eq:genact})
in 2-dimensional Minkowski spacetime.
Further consider the set of $M_n$ relations
in (\ref{eq:primconsgen}) for $n=1$
or (\ref{eq:rawconsgen}) for $n\geq2$. 
Suppose $M_n=2$ has been obtained,
with the relations
arranged into a column vector of the form
\begin{align}
\label{toycons1}
\boldsymbol{\phi}=\left(
\begin{array}{ccc}
F \\
F^\prime
\end{array}
\right),
\qquad
F=F(Q^A,\partial_\mu Q^A).
\end{align}
(For simplicity, we omit indices indicating the iteration.)
In view of the spatial derivatives' order difference between the two relations, a generic ansatz to (\ref{eq:nullfunct})
for $n=1$ or to (\ref{eq:nPhisnull}) for $n\geq2$
is particularly simple in this case:
\begin{align}
\label{eq:nullans1}
\boldsymbol{\Gamma}=
\big(\Gamma_0 +\Gamma_1\partial_1, \quad
\widetilde{\Gamma}_0 \big),
\end{align}
which readily yields a single linearly independent solution $m_n=1$ parametrized by $\Gamma_0=0$ and $\Gamma_1=-\widetilde{\Gamma}_0$.
We choose a convenient normalization for the solution,
look into its algebraic subspace
and choose a convenient normalization
for the $M_n-m_n=1$ linearly independent orthogonal vector:
\begin{align}
\label{eq:Gammabunch1}
\boldsymbol{\Gamma}=(-\partial_1,\, 1),
\qquad
\boldsymbol{\Gamma}_0=(0,\, 1),
\qquad
\boldsymbol{\Gamma}_0^\perp=(1,\,0).
\end{align}
The left contraction of the latter with (\ref{toycons1}) yields a functionally independent relation:
$\phi_\ast=\boldsymbol{\Gamma}_0^\perp\cdot \boldsymbol{\phi}=F$.
For $n=1$, $\phi_\ast$ can be regarded as the primary Lagrangian constraint.
For $n\geq2$, functional independence of $\phi_\ast$
with respect to Lagrangian constraints unveiled in previous iterations must be ensured before
regarding $\phi_\ast$ as the $n$-stage Lagrangian constraint.

\vspace*{0.5cm}

\noindent \textit{Toy model II}\\
Consider a theory of the form (\ref{eq:genact})
in 3-dimensional Minkowski spacetime.
Further, consider the following set of $M_n=3$ relations
in (\ref{eq:primconsgen}) for $n=1$
or (\ref{eq:rawconsgen}) for $n\geq2$,
arranged into a column vector 
\begin{align}
\label{toycons2}
\boldsymbol{\phi}=\left(
\begin{array}{ccc}
F_x \\
F+G_y \\
G
\end{array}
\right),
\qquad
F_x\equiv\partial_1 F,
\qquad 
G_y\equiv\partial_2 G,
\end{align}
where $(F,G)$ denote obviously functionally independent relations; for instance
$F=F(Q^A,\widehat{Q}^B,\partial_\mu Q^A)$
and $G=G(Q^A,\partial_\mu Q^A)$,
with the hat denoting a specific coordinate
within the set $Q^A$ that is not present.
Observation of the relative difference in the order of spatial derivatives between the relations leads us to postulate a generic ansatz to (\ref{eq:nullfunct})
for $n=1$ or to (\ref{eq:nPhisnull}) for $n\geq2$
of the form
\begin{align}
\label{eq:nullans2}
\boldsymbol{\Gamma}=
\big( 
\Gamma_0 +\Gamma^x_1\,\partial_1+\Gamma^y_1\,\partial_2, \quad
\widetilde{\Gamma}_0 +\widetilde{\Gamma}^x_1\,\partial_1+\widetilde{\Gamma}^y_1\,\partial_2,
\quad
\widehat{\Gamma}_0 +\widehat{\Gamma}^x_1\,\partial_1+\widehat{\Gamma}^y_1\,\partial_2+\widehat{\Gamma}^{xx}_2\,\partial_1\partial_1+2\widehat{\Gamma}^{xy}_2\,\partial_1\partial_2+\widehat{\Gamma}^{yy}_2\,\partial_2\partial_2
\big).
\end{align}
The above readily yields a single linearly independent solution $m_n=1$ parametrized by $ \widetilde{\Gamma}^x_1=-\Gamma_0$ and $\widehat{\Gamma}^{xy}_2=\Gamma_0/2 $,
with all other free functions set to zero.
We choose a convenient normalization for this solution,
look into its algebraic subspace
and choose a convenient normalization
for the $M_n-m_n=2$ linearly independent orthogonal vectors:
\begin{align}
\label{eq:Gammabunch2}
\boldsymbol{\Gamma}=\big(1,\,-\partial_1,\,\partial_{(1}\partial_{2)}\big),
\qquad
\boldsymbol{\Gamma}_0=(1,\, 0,\,0),
\qquad
\boldsymbol{\Gamma}_0^\perp=(0,\,1,\,0),
\qquad
\widetilde{\boldsymbol{\Gamma}}_0^\perp=(0,\,0,\,1).
\end{align}
The left contraction of the last two with (\ref{toycons2}) yields two functionally independent relations
\begin{align}
\label{eq:lagconstoy2}
\phi_\ast=
\boldsymbol{\Gamma}_0^\perp
\cdot\boldsymbol{\phi}=
F+G_y,
\qquad
\widetilde{\phi}_\ast=
\widetilde{\boldsymbol{\Gamma}}_0^\perp
\cdot\boldsymbol{\phi}=
G.
\end{align}
For $n=1$, (\ref{eq:lagconstoy2}) can be readily regarded as the primary Lagrangian constraints.
For $n\geq2$, functional independence of (\ref{eq:lagconstoy2})
with respect to Lagrangian constraints unveiled in previous iterations must be ensured before
reaching such a conclusion.

Notice that
the simplest choice of orthogonal vectors $(\boldsymbol{\Gamma}_0^\perp,
\widetilde{\boldsymbol{\Gamma}}_0^\perp)$
does not yield the obviously simplest
span of the constraint space,
given by $\{F,G\}$. 
Toy model II thus illustrates our first remark
in section (\ref{sec:rem}).

\subsection{Floreanini-Jackiw chiral boson}
\label{sec:chiral}

The Lagrangian for the 2-dimensional theory
of a chiral boson 
due to Floreanini and Jackiw~\cite{FJBoson87} is
\begin{align}
\label{eq:FJLag}
\mathcal{L}_{\textrm{FJ}}=
\frac{1}{2}\phi^\prime\left(
\dot{\phi}
-\phi^\prime\right).
\end{align}
Here, the scalar field $\phi=\phi(x^0,x^1)$ is
the only a priori independent field variable $Q^A$
and so $N=1$.
As is well-known,
this theory possesses no local symmetries
--- neither of the relevant form (\ref{eq:fieldtrans})
nor otherwise ---
in its original formulation (\ref{eq:FJLag}).
Therefore, $g,e=0$.
It is worth mentioning that a manifestly
Lorentz invariant action for
the Floreanini-Jackiw chiral boson exists, which has been further generalized into the so-called
$2k$-form electrodynamics
family of higher-dimensional theories~\cite{Townsend:2019koy}.

The (primary) equations of motion following from (\ref{eq:FJLag})
are of the form (\ref{eq:ELmatrix}),
with
\begin{align}
\label{eq:WUforFJ}
\mathbf{W}^{(1)}=0,
\qquad 
\mathbf{U}^{(1)}=
\dot{\phi}^\prime-\phi^{\prime\prime}.
\end{align} 
Obviously, the (row) rank of the (primary) Hessian is zero.
A convenient left null vector for it is simply
$\mathbf{V}^{(1)}=1$.
As per (\ref{eq:primconsgen})
and since no identical vanishing happens,
\begin{align}
\label{eq:primFJ}
\phi^{(1)}:=
\dot{\phi}^\prime-\phi^{\prime\prime}=0
\end{align}
can be readily regarded as
the only primary Lagrangian constraint
in the theory $l_1=1$.

The (primary) equations of motion,
together with the demand for stability
under time evolution
of the primary Lagrangian constraint,
conform the starting point of the second iteration (\ref{eq:secELs}),
where
\begin{align}
\label{eq:timeevolFJ}
\mathbf{W}^{(2)\downarrow}=
\left(
\begin{array}{ccc}
0 \\
\partial_1
\end{array}
\right),
\qquad
\mathbf{U}^{(2)\downarrow}=
\left(
\begin{array}{ccc}
\mathbf{U}^{(1)} \\
-\dot{\phi}^{\prime\prime}
\end{array}
\right).
\end{align} 
Clearly,
$\mathbf{W}^{(2)\downarrow}$
only admits left null vectors
of the form
\begin{align}
\label{eq:secnullFJ}
\mathbf{V}^{(2)\downarrow}=
(\mathbf{V}^{(1)},0).
\end{align}
According to the discussion below
(\ref{eq:gennullnit}),
there exists no secondary Lagrangian
constraint in the theory $l_2=0$.
The constraint algorithm thus terminates,
by means of closure \ref{item:Wclose}.

Using (\ref{eq:masterfor})
and (\ref{eq:lcount}),
we reproduce the renowned result
that the theory propagates
$N_{\textrm{DoF}}=1/2$ degrees of freedom.

\subsection{Maxwell electrodynamics}
\label{sec:maxwell}

The Lagrangian for
standard electrodynamics
in $d\geq2$ dimensions is
\begin{align}
\label{eq:MaxLag}
\mathcal{L}_{\textrm{M}}=
-\frac{1}{4}F_{\mu \nu}F^{\mu \nu},
\qquad
F_{\mu\nu}=2\partial_{[\mu}A_{\nu]},
\end{align}
where the components of
the vector field $A_{\mu}=A_{\mu}(x^0,x^i)$
conform the a priori independent
field variables $Q^A$
and thus $N=d$.
The theory enjoys
a manifest $U(1)$ gauge invariance,
under the transformation
\begin{align}
\label{eq:U1trans}
A_\mu \rightarrow A_\mu +\partial_\mu \theta,
\end{align}
which is of the relevant form (\ref{eq:fieldtrans}).
It follows that
$g=1$ and $e=2$.

The (primary) equations of motion
for $A_\mu$ following from (\ref{eq:MaxLag}) 
are of the form (\ref{eq:ELmatrix}),
with 
\begin{align}
\label{eq:primMax}
W^{(1)\mu\nu}=
\eta^{\mu\nu}
+\eta^{\mu0}\eta^{\nu 0}\equiv
\mathcal{W}^{\mu\nu},
\qquad
U^{(1)\mu}=
2\left(\eta^{0(\mu}\eta^{\nu)i}
\partial_i\dot{A}_\nu
+\eta^{i[\nu}\eta^{j]\mu}
\partial_i\partial_j A_\nu
\right)\equiv \mathcal{A}^\mu.
\end{align}
It is easy to see
that the (row) rank of the (primary) Hessian
is $d-1$.
A convenient left null vector for it
is $\mathbf{V}^{(1)}=\delta_\mu^0$.
As per (\ref{eq:primconsgen})
and since no identical vanishing happens,
\begin{align}
\label{eq:primconMax}
\phi^{(1)}:=\partial^i F_{i0}=0
\end{align}
can be identified with the primary Lagrangian constraint in the theory $l_1=1$.
In fact, this is Gauss's Law for the electric field.

The (primary) equations of motion,
together with the demand for stability
under time evolution
of the primary Lagrangian constraint,
conform the starting point of the second iteration.
They can be written as (\ref{eq:secELs}),
where
\begin{align}
\label{eq:secMax}
\mathbf{W}^{(2)\downarrow}=
\left(
\begin{array}{ccc}
\mathcal{W}^{\mu\nu} \\
-\eta^{i\nu}\partial_i
\end{array}
\right),
\qquad
\mathbf{U}^{(2)\downarrow}=
\left(
\begin{array}{ccc}
\mathcal{A}^\mu \\
\nabla^2 \dot{A}_0
\end{array}
\right).
\end{align}
It is easy to see that,
up to normalization,
there exists
only one linearly independent left null vector
to $\mathbf{W}^{(2)\downarrow}$
that does not trivially extend
$\mathbf{V}^{(1)}$.
We choose it as
$\mathbf{V}^{(2)\downarrow}=(\delta_\mu^i\partial_i,1)$.
We remark that the above is a particular instance of the general form prescribed in
(\ref{eq:gennullnit}).
As per (\ref{eq:rawconsgen}),
\begin{align}
\label{eq:secConMax}
\phi^{(2)}:=
\partial^i\partial^j F_{ij}
\equiv 0.
\end{align}
Hence, no secondary Lagrangian constraint exists
$l_2=0$.
The constraint algorithm thus terminates,
by means of closure \ref{item:trivialclose}.

Using (\ref{eq:masterfor})
and (\ref{eq:lcount}),
we obtain the familiar result
$N_{\textrm{DoF}}=d-2$.

\subsection{Proca electrodynamics}
\label{sec:proca}

Consider the simplest
massive electrodynamics theory
in $d\geq2$ dimensions
\begin{align}
\label{eq:ProcaLag}
\mathcal{L}_{\textrm{P}}=
-\frac{1}{4}F_{\mu \nu}F^{\mu \nu}
-\frac{1}{2}m^2A_\mu A^\mu,
\qquad
m\in\mathbb{R}_{>0}.
\end{align}
As for Maxwell electrodynamics before,
the components of
the vector field $A_{\mu}=A_{\mu}(x^0,x^i)$
conform the a priori independent
field variables $Q^A$
and thus $N=d$.
Contrastively,
the mass term explicitly breaks
the $U(1)$ gauge invariance
of (\ref{eq:MaxLag}),
leaving no residual symmetry.
Hence,
$g=e=0$.

The (primary) equations of motion for $A_\mu$
following from (\ref{eq:ProcaLag}) are of the form (\ref{eq:ELmatrix}), with 
\begin{align}
\label{eq:WUforProcaJ}
W^{(1)\mu\nu}=
\mathcal{W}^{\mu\nu},
\qquad
U^{(1)\mu}=
\mathcal{A}^\mu+m^2A^\mu,
\end{align}
where $(\mathcal{W}^{\mu\nu},\mathcal{A}^\mu)$ were defined in (\ref{eq:primMax}).
The (primary) Hessian
is the same as for Maxwell electrodynamics
(\ref{eq:primMax}), with rank $d-1$.
We again chose $\mathbf{V}^{(1)}=\delta_\mu^0$
as a convenient left null vector for it.
Using (\ref{eq:primconsgen})
and since no identical vanishing happens,
\begin{align}
\label{eq:primconProca}
\phi^{(1)}:=
\partial^i F_{i0}-m^2A_0=0
\end{align}
can be identified with the primary Lagrangian constraint in the theory $l_1=1$.

Next, consider
the (primary) equations of motion,
together with the demand for stability
under time evolution
of the primary Lagrangian constraint,
in the form (\ref{eq:secELs}),
where
\begin{align}
\label{eq:secPro}
\mathbf{W}^{(2)\downarrow}=
\left(
\begin{array}{ccc}
\mathcal{W}^{\mu\nu} \\
-\eta^{i\nu}\partial_i
\end{array}
\right),
\qquad
\mathbf{U}^{(2)\downarrow}=
\left(
\begin{array}{ccc}
\mathcal{A}^\mu+m^2A^\mu \\
(\nabla^2 
-m^2)\dot{A}_0
\end{array}
\right).
\end{align}
The above conforms the starting point of
the second iteration
in the constraint algorithm.
We note that $\mathbf{W}^{(2)\downarrow}$
for Proca electrodynamics
matches that of Maxwell's theory
(\ref{eq:secMax})
and
we repeat our choice
$\mathbf{V}^{(2)\downarrow}=(\delta^i_\mu\partial_i,1)$
for a conveniently normalized 
left null vector
that is linearly independent from
a trivial extension of $\mathbf{V}^{(1)}$.
Using (\ref{eq:rawconsgen}), we obtain
the relation
\begin{align}
\label{eq:secConProca}
\phi^{(2)}:=
m^2\partial^\mu A_\mu=0.
\end{align}
Clearly, the above does not identically vanish.
It is also easy to see that 
(\ref{eq:primconProca}) and (\ref{eq:secConProca}) are functionally independent.
Therefore, $l_2=1$
and (\ref{eq:secConProca})
can be taken
as the secondary Lagrangian constraint.

In a third iteration of the constraint algorithm,
we
consider
the (primary) equations of motion,
along with the demand for stability
under time evolution
of both the primary and the secondary Lagrangian constraints,
in the form (\ref{eq:secELs}),
where the tertiary equations of motion
are given by
\begin{align}
\label{eq:tertPro}
\mathbf{W}^{(3)}=
\left(m^2\eta^{\mu 0}\right),
\qquad
\mathbf{U}^{(3)}=\big(m^2
\nabla\dot{A}\big).
\end{align}
It is obvious that
$\mathbf{W}^{(3)\downarrow}$
does not admit left null vectors
beyond trivial extensions of
$(\mathbf{V}^{(1)},\mathbf{V}^{(2)})$
before.
Consequently,
there exists no tertiary Lagrangian constraints $l_3=0$ and
the algorithm terminates according to closure \ref{item:Wclose}.

Using (\ref{eq:masterfor}) and (\ref{eq:lcount}), we reproduce the well-known result that Proca electrodynamics propagates $N_{\textrm{DoF}}=d-1$ degrees of freedom.

\subsection{Podolsky electrodynamics}
\label{sec:podol}

Podolsky's proposal for a generalized
electrodynamics theory~\cite{Podolsky:1942zz}
\begin{align}
\label{eq:podolor}
\mathcal{L}_{\textrm{Po}}=
-\frac{1}{4}F_{\mu\nu}F^{\mu\nu}
-\frac{a^2}{2}\partial_\mu F^{\mu\nu}
\partial^\rho F_{\rho\nu}
\end{align}
is arguably the best-known higher-order field theory. So as to remain within the scope of section \ref{sec:method},
we consider its first-order formulation~\cite{Thibes:2016ivt} in
$d\geq2$ dimensions
\begin{align}
\label{eq:LThibes}
\mathcal{L}_{\textrm{Po1}}=
-\frac{1}{4}F_{\mu \nu}F^{\mu \nu}
+\frac{a^2}{2}B_{\mu}B^{\mu}
-\frac{a^2}{2}G_{\mu\nu}F^{\mu \nu},
\qquad
G_{\mu\nu}=2\partial_{[\mu}B_{\nu]},
\qquad
a\in\mathbb{R}_{>0},
\end{align}
The components of the vector fields $A_{\mu}=A_{\mu}(x^0,x^i)$, $B_{\mu}=B_{\mu}(x^0,x^i)$ are the a priori independent field variables $Q^A$ and hence $N=2d$. The Lagrangian (\ref{eq:LThibes}) inherits the symmetry of Maxwell electrodynamics and is gauge invariant under the field transformations
\begin{align}
\label{eq:SymmPodolsky}
A_\mu \rightarrow A_\mu +\partial_\mu \theta, \qquad
B_\mu \rightarrow B_\mu,
\end{align}
which are of the relevant form (\ref{eq:fieldtrans}).
It follows that
$g=1$ and $e=2$, as in the Maxwell case earlier on.

The (primary) equations of motion following from (\ref{eq:LThibes}) are of the form (\ref{eq:ELmatrix}), with 
\begin{align}
\label{eq:WUforPodolskyJ}
\mathbf{W}^{(1)}=
\begin{pmatrix}
1 & a^2  \\
a^2 & 0 
\end{pmatrix}
\mathcal{W}^{\mu\nu},
\qquad
\mathbf{U}^{(1)}=
\begin{pmatrix}
\mathcal{A}^\mu
+a^2
\mathcal{B}^\mu
\vspace*{0.2cm} \\
a^2\mathcal{A}^\mu
-a^2B^\mu
\end{pmatrix},
\end{align}
where $(\mathcal{W}^{\mu\nu}, \mathcal{A}^\mu)$ were introduced in (\ref{eq:primMax})
and $\mathcal{B}^\mu$ stands for the same quantity
as $\mathcal{A}^\mu$, but in terms of $B^\mu$ instead of $A^\mu$.
It is rather obvious that the (row) rank
of the (primary) Hessian is $2(d-1)$. A convenient choice for the two linearly independent left null vectors is $\mathbf{V}_1^{(1)}=(\delta^0_{\mu}, 0)$ and $\mathbf{V}_2^{(1)}=(0, \delta^0_{\mu})$.
Using (\ref{eq:primconsgen}), since no identical vanishing happens and taking into consideration
the manifest functional independence,
\begin{align}
\label{eq:primconPodolsky}
\boldsymbol{\Phi}^{(1)}:=
\left(
\begin{array}{ccc}
\partial^i F_{i0} + a^2 \partial^iG_{i0} \vspace*{0.2cm}\\
a^2 \partial^i F_{i0} +a^2 B_0
\end{array}
\right)=0
\end{align}
can be identified with the two primary Lagrangian constraints in the theory $l_1=2$.

In order to ensure the stability of the primary Lagrangian constraints, we calculate the theory's secondary equations of motions (\ref{eq:EL2matrix}).
We find
\begin{align}
\label{eq:secEoMPod}
\mathbf{W}^{(2)}=
\begin{pmatrix}
-1 & -a^2 \\
-a^2 & 0
\end{pmatrix}
\eta^{\mu i}\partial_i,
\qquad 
\mathbf{U}^{(2)}=
\begin{pmatrix}
\nabla^2  \dot{A}_0 + a^2\nabla^2 \dot{B}_0
\vspace*{0.2cm}\\
a^2\nabla^2 \dot{A}_0 +a^2 \dot{B}_0
\end{pmatrix}
\end{align}
and consider them
together with the (primary) equations of motion,
as described in (\ref{eq:secELs}) and (\ref{eq:secHessU}).
Beyond trivial extensions of $\mathbf{V}^{(1)}_1$ and $\mathbf{V}^{(1)}_2$ before and up to normalization,
there exists another linearly independent
left null vector $\mathbf{V}^{(2)\downarrow}$ to $\mathbf{W}^{(2)\downarrow}$.
It can be found as prescribed around (\ref{eq:nnullhess}). Explicitly,
we postulate
\begin{align}
\label{eq:null2ansatz}
\mathbf{V}^{(2)\downarrow}=
\big(V_\mu^i\,\partial_i,\quad
\widetilde{V}_\mu^i\,\partial_i,\quad
\widehat{V},\quad
\bar{V}\big)
\quad
\textrm{such that } \,
\begin{cases}
(V_\mu^i+a^2\widetilde{V}_\mu^i)\,
\mathcal{W}^{\mu\nu}
-(\widehat{V}+a^2\bar{V})\,\eta^{\nu i}=0,
\vspace*{0.2cm}\\
V_\mu^i\,
\mathcal{W}^{\mu\nu}
-\widehat{V}\,\eta^{\nu i}=0.
\end{cases}
\end{align}
Our simplified ansatz is a direct consequence of 
the overall spatial derivatives' order difference between $\mathbf{W}^{(1)}$ and $\mathbf{W}^{(2)}$.
We choose as representative of
the one-parameter family of linearly independent solutions 
\begin{align}
\label{eq:null2inpod}
\mathbf{V}^{(2)\downarrow}=
\big(a^2\delta^i_{\mu}\,\partial_i, \quad
-\delta^i_{\mu}\,\partial_i, \quad
a^2, \quad -1
\big).
\end{align}
Using the above and (\ref{eq:rawconsgen}), we obtain the relation
\begin{align}
\label{eq:SecConsPodolsky}
\phi^{(2)}:=a^2\partial^{\mu} B_{\mu}=0.
\end{align}
It is easy to see that (\ref{eq:SecConsPodolsky}) does neither identically vanish nor functionally depend on the primary Lagrangian constraints (\ref{eq:primconPodolsky}). Therefore, $l_2=1$ and (\ref{eq:SecConsPodolsky}) can be regarded as the secondary Lagrangian constraint in the theory.
Note it matches the secondary Lagrangian constraint (\ref{eq:secConProca}) in Proca electrodynamics.

In a third iteration of the constraint algorithm, we consider the (primary) equations of motion, together with the demand for stability of the primary and secondary Lagrangian constraints, in the form (\ref{eq:ELnarrow}).
For the tertiary equations of motion,
we obtain
\begin{align}
\label{eq:tertPodolsky}
\mathbf{W}^{(3)}=
\begin{pmatrix}
0, & a^2\eta^{\mu 0} 
\end{pmatrix},
\qquad 
\mathbf{U}^{(3)}=\big(a^2\nabla\dot{B}\big).
\end{align}
It is a matter of conjunct inspection of
$\big(\mathbf{W}^{(1)},\mathbf{W}^{(2)},\mathbf{W}^{(3)}\big)$ to deduce that
$\mathbf{W}^{(3)\downarrow}$ admits no other left null vectors beyond trivial extensions of the left null vectors obtained in earlier iterations. Consequently, no tertiary Lagrangian constraints exist $l_3=0$ and the algorithm terminates according to closure \ref{item:Wclose}.

Using (\ref{eq:masterfor}) and (\ref{eq:lcount}), we reproduce the well-known result $N_{\textrm{DoF}}=2d-3$.
This count supports the interpretation of Podolsky electrodynamics
as the theory of an interacting pair of vector fields, one of which is massive e.g.~\cite{Galvao:1986yq,Fonseca:2010av}.
It is also instructive to compare the constraint structures in this and the two previous sections \ref{sec:maxwell} and \ref{sec:proca}.

\subsection{Minimal Model in Extended Proca-Nuevo}
\label{sec:MMEPN}

We consider
the simplest, 2-dimensional case in the so-called Extended Proca-Nuevo (EPN)
class of massive vector field theories,
dubbed Minimal Model.
EPN was originally proposed
in~\cite{deRham:2021efp},
as an extension of the Proca-Nuevo (PN) construction in~\cite{deRham:2020yet}.
A study of the constraint structure of both PN and EPN,
including an explicit discussion of the Minimal Model here revisited,
can be found
in~\cite{deRham:2023brw}.

The Lagrangian for the Minimal Model is
\begin{align}
\label{eq:LagMMPN}
\mathcal{L}_{\textrm{MM}}=
\Lambda^2(\alpha +2 N-4),
\end{align}
where $\Lambda$ is a constant of length dimension $(-1)$,
$\alpha$ is a dimensionless and at least twice differentiable function
of the square of the vector
\begin{align}
\label{eq:alphadef}
\alpha=
\alpha(X),
\qquad
X=-A_0^2+A_1^2,
\end{align}
and $N\neq 0$ is given by
\begin{align}
\label{eq:varsdef}
N=\sqrt{x^2-y^2},
\qquad
x=2+\frac{A_1^\prime-\dot{A}_0}{\Lambda},
\qquad
y=\frac{\dot{A}_1-A_0^\prime}{\Lambda}.
\end{align}
In (\ref{eq:LagMMPN}),
the components of the vector field $A_\mu=A_\mu(x^0,x^1)$
are the a priori independent field variables $Q^A$ and so $N=2$.
By definition, theories within the (E)PN
class(es) do not possess any local symmetry.
Therefore, $g,e=0$.

The (primary) equations of motion following from
(\ref{eq:LagMMPN}) are of the form
(\ref{eq:ELmatrix}), with
\begin{align}
\label{eq:WUMMprim}
\mathbf{W}^{(1)}=
-\frac{2}{N^3}
\left(
\begin{array}{ccc}
y^2 & xy \\
xy & x^2
\end{array}
\right),
\qquad
\mathbf{U}^{(1)}=
\left(
\begin{array}{cccc}
\displaystyle
\frac{2}{N^3}\left[xy(2\dot{A}_0^\prime-A_1^{\prime\prime})
-x^2A_0^{\prime\prime}+\bar{N}^2\dot{A}_1^\prime\right]
+2\Lambda^2\alpha_XA_0 \vspace*{0.2cm} \\
\displaystyle
\frac{2}{N^3}\left[xy(2\dot{A}_1^\prime-A_0^{\prime\prime})
-y^2A_1^{\prime\prime}+\bar{N}^2\dot{A}_0^\prime\right]
-2\Lambda^2\alpha_XA_1
\end{array}
\right),
\end{align}
where we have introduced the short-hands
\begin{align}
\label{eq:newvars}
\bar{N}=
\sqrt{x^2+y^2},
\qquad 
\alpha_X=
\frac{d\alpha (X)}{dX},
\qquad 
\alpha_{XX}=
\frac{d^2\alpha (X)}{dX^2},
\end{align}
and the second (total) derivative of $\alpha$ has been defined
for later convenience.
It is easy to see that the (row) rank of the (primary) Hessian is 1.
Following~\cite{deRham:2020yet,deRham:2023brw},
we choose a left null vector to $\mathbf{W}^{(1)}$ as
\begin{align}
\mathbf{V}^{(1)}=\frac{1}{N}(x,-y).
\end{align}
As per (\ref{eq:primconsgen})
and since no identical vanishing happens,
\begin{align}
\label{eq:primMM}
\phi^{(1)}:=
\frac{2\Lambda}{N^2}(xy^\prime-yx^\prime)
+\frac{2\Lambda^2}{N}\alpha_X(xA_0+yA_1)
=0
\end{align}
can be readily regarded as
the only primary Lagrangian constraint
in the theory $l_1=1$.

Following (\ref{eq:EL2matrix}),
at the second iteration we find
\begin{align}
\label{eq:wuexpl}
\begin{array}{lll}
\mathbf{W}^{(2)}=&\hspace*{-0.2cm}
\big(
\begin{array}{ccc}
\Omega_1 & \Omega_2 
\end{array}
\big), \vspace*{0.2cm} \\
\mathbf{U}^{(2)}\hspace*{0.1cm}=&\hspace*{-0.2cm}
\displaystyle
-\frac{2}{N^2}
(x\dot{A}_0^{\prime\prime}+y\dot{A}_1^{\prime\prime})
+\frac{2\overline{N}^2}{N^4}
(x^\prime \dot{A}_0^\prime-y^\prime \dot{A}_1^\prime)
-\frac{4xy}{N^4}
(y^\prime \dot{A}_0^\prime-x^\prime \dot{A}_1^\prime)
\vspace*{0.3cm} \\
&\hspace*{-0.2cm}
\displaystyle
-\frac{2\Lambda}{N^3}\alpha_X
(yA_0+xA_1)
(x \dot{A}_0^\prime+y \dot{A}_1^\prime)
+\frac{2\Lambda^2}{N}\alpha_X
(x \dot{A}_0+y \dot{A}_1)
+\frac{2\Lambda^2}{N}\alpha_{XX}
(xA_0+yA_1)
\dot{X},
\end{array}
\end{align}
where we have introduced
\begin{align}
\label{eq:dotpieces}
\begin{array}{llll}
\Omega_1=&\hspace*{-0.2cm}
\displaystyle
\omega_1 + \overline{\omega}_1\partial_1=
\frac{4x}{N^4}(xy^\prime-yx^\prime)
-\frac{2y^\prime}{N^2}
+\frac{2\Lambda}{N^3}\alpha_X y(xA_1+yA_0)
+\frac{2y}{N^2}\partial_1,
\vspace*{0.3cm} \\
\Omega_2=&\hspace*{-0.2cm}
\displaystyle
\omega_2 + \overline{\omega}_2\partial_1=
\frac{4y}{N^4}(xy^\prime-yx^\prime)
-\frac{2x^\prime}{N^2}
+\frac{2\Lambda}{N^3}\alpha_X x(xA_1+yA_0)
+\frac{2x}{N^2}\partial_1.
\end{array}
\end{align}
As per (\ref{eq:secELs})
and (\ref{eq:secHessU}), we proceed to the
joint consideration of the primary and secondary equations of motion.
In particular, we 
inspect the row rank of $\mathbf{W}^{(2)\downarrow}$ via a conveniently normalized
generic left null vector of the form (\ref{eq:gennullnit})
\begin{align}
\label{eq:secnull}
\mathbf{V}^{(2)\downarrow}=(V_0+V_1\partial_1,\,
\widetilde{V}_0+\widetilde{V}_1\partial_1,\,1).
\end{align}
For the above to fulfill (\ref{eq:nnullhess}),
four a priori independent equations must be fulfilled.
These are
\begin{align}
\label{eq:secnullconds}
\begin{cases}
\displaystyle
-\frac{2y}{N^3}(V_1y+\widetilde{V}_1x)
+\overline{\omega}_1=0,
\vspace*{0.3cm} \\
\displaystyle
-\frac{2x}{N^3}(V_1y+\widetilde{V}_1x)
+\overline{\omega}_2=0, 
\vspace*{0.3cm} \\
\displaystyle
-\frac{2y}{N^3}(V_0y+\widetilde{V}_0x)
-\frac{2y}{N^3}(V_1y^\prime+\widetilde{V}_1x^\prime)
-2\left(\partial_1\frac{y}{N^3}\right)(V_1y+\widetilde{V}_1x)
+\omega_1=0,
\vspace*{0.3cm} \\
\displaystyle
-\frac{2x}{N^3}(V_0y+\widetilde{V}_0x)
-\frac{2x}{N^3}(V_1y^\prime+\widetilde{V}_1x^\prime)
-2\left(\partial_1\frac{x}{N^3}\right)(V_1y+\widetilde{V}_1x)
+\omega_2=0.
\end{cases}
\end{align}
The first (second) equation can be easily seen to be redundant
with respect to the second (first) equation.
Assuming $x,y\neq0$,
either of these equations poses
the same null vector condition on (\ref{eq:secnull}):
\begin{align}
\label{eq:nullcond1}
\frac{1}{N}(V_1y+\widetilde{V}_1x)-1=0.
\end{align}
Additionally and
using (\ref{eq:nullcond1}),
the third and fourth equations
can be seen to boil down
to the same null vector condition on (\ref{eq:secnull}):
\begin{align}
\label{eq:nullcond2}
V_0y+\widetilde{V}_0x
+\frac{xy^\prime-yx^\prime}{y}
\left(\frac{x}{N}-\widetilde{V}_1\right)
-\Lambda\alpha_X(xA_1+yA_0)
=0.
\end{align}
On the whole, a (unique, up to normalization) linearly independent solution is given by
\begin{align}
V_0=\Lambda \alpha_X A_0,
\qquad
\widetilde{V}_0=-\frac{y}{N},
\qquad,
V_1=\Lambda\alpha_XA_1,
\qquad
\widetilde{V}_1=\frac{x}{N}.
\end{align}
Pursuing the constraint algorithm
by means of (\ref{eq:rawconsgen}),
we find that 
the left contraction of the above null vector with
$\mathbf{E}^{(2)\downarrow}$
gives rise to the relation
\begin{align}
\label{eq:secconsMM}
\begin{array}{llll}
\phi^{(2)}:=& \hspace*{-0.2cm}
\displaystyle
\frac{2\Lambda}{N^4}
(xy^\prime-yx^\prime)^2
+\frac{2\Lambda^3}{N}\alpha_X
(2x-N^2)
+\frac{2\Lambda^2}{N^3}\alpha_X
(xy^\prime-yx^\prime)(xA_0+yA_1)
-2\Lambda^3\alpha_X^2 X
\vspace*{0.3cm} \\
& \hspace*{-0.2cm} \displaystyle
+\frac{2\Lambda^2}{N}\alpha_{XX}\left[
\dot{X}(xA_0+yA_1)-X^\prime(yA_0+xA_1)
\right]=0.
\end{array}
\end{align}
Close inspection readily reveals
that (\ref{eq:primMM}) and (\ref{eq:secconsMM})
are functionally independent.
Complementarily,
it can be checked that there exists no non-trivial solution to (\ref{eq:nullvarphi})
for the Minimal Model. 
As a result, there exists a secondary constraint in the theory $l_2=1$, which can be parametrized by (\ref{eq:secconsMM}) itself.
However, it is more convenient
to use (\ref{eq:primMM}) so as to
simplify (\ref{eq:secconsMM}) to
\begin{align}
\label{eq:weaksecconsMM}
\phi^{(2)}=
\frac{2\Lambda^3}{N}\alpha_X
(2x-N^2)
-
2\Lambda^3\alpha_X^2 X
+\frac{2\Lambda^2}{N}\alpha_{XX}\left[
\dot{X}(xA_0+yA_1)-X^\prime(yA_0+xA_1)
\right]=0.
\end{align}

A posteriori, we streamline the discussion
on the third iteration for the Minimal Model.
In particular, we limit ourselves to proving that
$\mathbf{W}^{(3)\downarrow}$, as defined around
(\ref{eq:ELnarrow}), has maximal
row rank equal to 2. 
As a direct consequence,
there exist no tertiary Lagrangian constraints
$l_3=0$ and
the constraint algorithm thus terminates,
by means of closure \ref{item:Wclose}.

Consider the first and fourth row in
$\mathbf{W}^{(3)\downarrow}$.
All entries are scalars and thus it is
a matter of easy algebra to verify that
their determinant does not vanish:
\begin{align}
\label{eq:detnotzero}
\textrm{det}\left(
\begin{array}{ccc}
\displaystyle
-\frac{2y^2}{N^3} &
\displaystyle
-\frac{2xy}{N^3}
\vspace*{0.3cm}\\
\displaystyle
\frac{\partial\phi^{(2)}}
{\partial \dot{A}_0} &
\displaystyle
\frac{\partial\phi^{(2)}}
{\partial \dot{A}_1} 
\end{array}
\right)=\frac{4\Lambda^2 }{N^4}y
\left[
N^2\alpha_X
-2\alpha_{XX}(xA_0+yA_1)^2
\right]\neq 0.
\end{align}
Moreover, straightforward examination
unequivocally indicates that the above determinant is functionally independent
from both the primary (\ref{eq:primMM}) and the secondary (\ref{eq:weaksecconsMM}) Lagrangian constraints.
This proves there exists a maximal row rank
minor within $\mathbf{W}^{(3)\downarrow}$.
Hence, $\mathbf{W}^{(3)\downarrow}$ itself
has maximal row rank
and does not admit left null vectors
that non-trivially extend $\mathbf{V}^{(1)}$
and $\mathbf{V}^{(2)\downarrow}$ above.

Using (\ref{eq:masterfor}) and (\ref{eq:lcount}), we independently reproduce the recent result in~\cite{deRham:2023brw} that the Minimal Model propagates $N_{\textrm{DoF}}=1$ degree of freedom,
as a massive electrodynamics theory in 2 dimensions must do.

\subsection{2-dimensional Palatini}
\label{sec:palatini}

Consider Einstein-Hilbert
theory of gravity in 2 dimensions.
A popular first-order reformulation
attributed to Palatini~\cite{palatini1919deduzione}
is 
\begin{align}
\label{eq:PalatiniL}
\mathcal{L}_{\textrm{Pa}}=-(\partial_\rho h^{\mu\nu})G^\rho_{\mu\nu}+h^{\mu\nu}(G^\rho_{\rho\mu}G^{\sigma}_{\sigma\nu}
-G^{\rho}_{\sigma\mu}G^{\sigma}_{\rho\nu}),
\end{align}
where $h^{\mu\nu}=h^{\nu\mu}$ and $G^{\lambda}_{\mu\nu}=G^{\lambda}_{\nu\mu}$ denote tensors proportional to the metric and the connection,
respectively.
Their independent components conform
the a priori independent field variables
\begin{align}
Q^A=
\{h\equiv h^{00},\, h^1\equiv h^{01},\, h^{11},\, G\equiv G^{0}_{00}, \,G_1\equiv  G^{0}_{10}, \,
G_{11}\equiv G^{0}_{11},\,\mathcal{G}^1\equiv G^{1}_{00},\, \mathcal{G}^1_1\equiv G^{1}_{10},\,\mathcal{G}^{1}_{11}\equiv G^{1}_{11}\}, 
\end{align}
where we have introduced multiple renamings for notational simplicity. It readily follows that $N=9$.
The theory is invariant under the field transformations~\cite{Kiriushcheva:2005sk,McKeon:2016vuq}
\begin{align}
\label{eq:SymmPalatini}
h^{\mu \nu} \rightarrow h^{\mu \nu}
+2\epsilon^{\rho(\mu}h^{\nu)\sigma}
\theta_{\rho\sigma},
\qquad
G^{\rho}_{\mu \nu} \rightarrow G^{\rho}_{\mu \nu} +\epsilon^{\rho\sigma}
\partial_\sigma\theta_{\mu\nu}
+2\epsilon^{\sigma\lambda}
G^\rho_{\sigma(\mu}\eta_{\nu)\lambda},
\end{align}
which are of the relevant form
(\ref{eq:fieldtrans}).
Here, $\theta_{\mu\nu}=\theta_{\nu\mu}$,
so it has $g=3$ independent components.
It is easy to see that $e=6$. 

As a direct consequence of the exclusively linear dependence of the Lagrangian \eqref{eq:PalatiniL} on the generalized velocities $\dot{Q}^A$,
the (primary) Hessian vanishes $\bm{W}^{(1)}=0$
and its linearly independent left null vectors can be chosen as
\begin{equation}
\label{eq:NV1_Pa}
(V^{(1)}_{I})^A=\delta^{A}_{I}, 
\qquad
I=1,2,\ldots,M_1=9.
\end{equation}
Then, the components of the $\bm{U}^{(1)}$ vector
coincide with the relations \eqref{eq:primconsgen}:
$\bm{U}^{(1)}=\boldsymbol{\phi}^{(1)}$,
where
\begin{align}
\displaystyle
\label{eq:primcon_Pa}
\hspace*{-0.5cm}
\begin{array}{llllllllll}
& \phi_{1}&\hspace*{-0.2cm}:=&\hspace*{-0.2cm}
-\big[\dot{G}+\partial_{1}\mathcal{G}^{1}+2(G \mathcal{G}^{1}_{1}-G_{1}\mathcal{G}^{1})\big]=0, 
&\qquad \quad
& \phi_{2}&\hspace*{-0.2cm}:=&\hspace*{-0.2cm}
-2\big(\dot{G}_{1}+\partial_{1}\mathcal{G}^{1}_{1}+G\mathcal{G}^{1}_{11}-G_{11}\mathcal{G}^{1}\big)=0,
\vspace*{0.2cm} \\
&\phi_{3}&\hspace*{-0.2cm}:=&\hspace*{-0.2cm}
-\big[\dot{G}_{11}+\partial_{1}\mathcal{G}^{1}_{11}+2(G_{1}\mathcal{G}^{1}_{11}-G_{11}\mathcal{G}^{1}_{1})\big]=0,
&\qquad \quad
&\phi_{4}&\hspace*{-0.2cm}:=&\hspace*{-0.2cm}
\dot{h}-2(h\mathcal{G}^{1}_{1}+h^1\mathcal{G}^{1}_{11})=0,
\vspace*{0.2cm}\\
&\phi_{5}&\hspace*{-0.2cm}:=&\hspace*{-0.2cm}
2\big( \dot{h}^{1}+h\mathcal{G}^{1}-h^{11}\mathcal{G}^{1}_{11}\big)=0, 
&\qquad \quad
&\phi_{6}&\hspace*{-0.2cm}:=&\hspace*{-0.2cm}
\dot{h}^{11}+2(h^1\mathcal{G}^{1}+h^{11}\mathcal{G}^{1}_{1})=0,
\vspace*{0.2cm} \\
&\phi_{7}&\hspace*{-0.2cm}:=&\hspace*{-0.2cm}
\partial_{1}h+2(hG_{1}+h^1G_{11})=0, 
&\qquad \quad
&\phi_{8}&\hspace*{-0.2cm}:=&\hspace*{-0.2cm}
2\big(\partial_{1}h^{1}-hG+h^{11}G_{11} \big)=0,
\vspace*{0.2cm} \\
&\phi_{9}&\hspace*{-0.2cm}:=&\hspace*{-0.2cm}
\partial_{1}h^{11}-2(h^1G+h^{11}G_{1})=0.
\end{array}
\end{align}
(For simplicity, we omit indices indicating the iteration.)
Observe that the first six relations depend on a distinct generalized velocity each,
while the latter three do not depend on generalized velocities.
Namely, functional independence is manifest
for $\{\phi_1,\ldots,\phi_6\}$.
Functional (in)dependence among $\phi_{\textrm{red}}=\{\phi_{7},\phi_8,\phi_9\}$ is nonobvious
and so we proceed to test it via
(\ref{eq:nullfunct}).
Postulating
\begin{align}
\label{eq:rednull}
\boldsymbol{\Gamma}_{\textrm{red}}=
\big(
\Gamma_0+\Gamma_1\,\partial_1, \quad
\widetilde{\Gamma}_0+
\widetilde{\Gamma}_1\,\partial_1, \quad
\widehat{\Gamma}_0+
\widehat{\Gamma}_1\partial_1
\big),
\end{align}
the reduced left null vector condition
$\boldsymbol{\Gamma}_{\textrm{red}}\cdot 
\boldsymbol{\phi}_{\textrm{red}}=0$
yields no non-trivial solution, without much algebraic effort.
It follows that ${\phi}_{\textrm{red}}$ indeed comprises functionally independent relations.
Consequently, $l_1=9$ and the relations \eqref{eq:primcon_Pa} themselves can be regarded as the primary Lagrangian constraints in the theory.

We proceed to the second iteration.
As per \eqref{eq:primtime}, the secondary equations of motion are given by
\begin{align}
\label{eq:W2st_Pa}
\bm{W}^{(2)}=
\left(
\begin{array}{ccccccccc}
0 & -\bm{w} & 0\\
\bm{w} & 0 & 0\\
0 & 0 & 0
\end{array}
\right),
\qquad
\bm{w}=\textrm{diag}(1,2,1),
\qquad
\bm{U}^{(2)}=
\left(
\begin{array}{ccccccccc}
\bm{U} \\
\dot{\phi}_{\text{red}} 
\end{array}
\right)
\end{align}
where the components of $\bm{U}$ are 
 \begin{align}
\displaystyle
\label{eq:Ucompts_Pa}
\hspace*{-0.5cm}
\begin{array}{llllllllll}
& U_{1}&\hspace*{-0.2cm}=&\hspace*{-0.2cm}
-\partial_{1}\dot{\mathcal{G}}^{1} -2 \,\partial_{0}\left({G}\,\mathcal{G}^{1}_{1} -G_{1}\mathcal{G}^{1}\right), 
&\qquad \quad
&U_{2}&\hspace*{-0.2cm}=&\hspace*{-0.2cm}
-2\,\partial_{1}\dot{\mathcal{G}}^{1}_{1} -2\, \partial_{0}\left({G}\,\mathcal{G}^{1}_{11} -G_{11}\mathcal{G}^{1}\right),
\vspace*{0.2cm} \\
&U_{3}&\hspace*{-0.2cm}=&\hspace*{-0.2cm}
-\partial_{1}\dot{\mathcal{G}}^{1}_{11} -2\, \partial_{0}\left(G_{1}\mathcal{G}^{1}_{11}-{G}_{11}\,\mathcal{G}^{1}_{1}\right),
&\qquad \quad
&U_{4}&\hspace*{-0.2cm}=&\hspace*{-0.2cm}
- 2\,\partial_{0} \left(  h\,\mathcal{G}^{1}_{1} + h^{1}\mathcal{G}^1_{11}\right),
\vspace*{0.2cm} \\
&U_{5}&\hspace*{-0.2cm}=&\hspace*{-0.2cm}
 2\,\partial_{0} \left(  h\,\mathcal{G}^{1} -h^{11}\,\mathcal{G}^{1}_{11}\right),
&\qquad \quad
&U_{6}&\hspace*{-0.2cm}=&\hspace*{-0.2cm}
2\partial_{0}\left(  h^{11}\,\mathcal{G}^{1}+h^{1}\mathcal{G}^{1}_{1}  \right).
\end{array}
\end{align}
It is evident that $\text{rrank}\left(\bm{W}^{(2)} \right)=6$.
Following (\ref{eq:nnullhess}) and leaving
aside trivial extensions of the primary left null vectors \eqref{eq:PalatiniL}, we find $M_2=3$ additional linearly independent left null vectors,
up to normalization.
We conveniently choose them as
\begin{equation}
\label{eq:secNV_Pa}
\mathbf{V}^{(2)\downarrow}_R=
\big(\underbrace{0,0,\dots,0}_{15\textrm{ components}},\hat{R}
\big),
\qquad 
R=1,2,3,
\end{equation}
where $\hat{R}$ denotes the standard Cartesian unit vector in Euclidean 3 dimensions.
In accordance with (\ref{eq:rawconsgen}),
we find the relations
\begin{align}
\label{eq:SCts_PA}
\begin{array}{llll}
\phi_{1}^{(2)}&\hspace*{-0.2cm}:=&\hspace*{-0.2cm}
\partial_1 \dot{h}+2(\dot{h}G_{1}+\dot{h}^1G_{11}+h\dot{G}_{1}+h^{1}\dot{G}_{11})=0, 
\vspace*{0.2cm} \\
\phi_{2}^{(2)}&\hspace*{-0.2cm}:=&\hspace*{-0.2cm} 2(\partial_1\dot{h}^{1}-\dot{h}G+\dot{h}^{11}G_{11}-h\dot{G}+h^{11}\dot{G}_{11})=0,  \vspace*{0.2cm}\\ 
\phi_{3}^{(2)}&\hspace*{-0.2cm}:=&\hspace*{-0.2cm}
\partial_1\dot{h}^{11}-2(\dot{h}^{1}G+\dot{h}^{11}G_1+h^{1}\dot{G}+h^{11}\dot{G}_{1})=0.
\end{array}
\end{align}
Following the prescription in \eqref{eq:nPhisnull}, the search for functional (in)dependence within the secondary stage involves an ansatz for a vector
$\boldsymbol{\Gamma}^{(2)}$
with up to two spatial derivatives.
Nonetheless, due to the relative structure of the first term in each relation, it is easy to see that $\boldsymbol{\Gamma}^{(2)}\cdot \bm{\phi}^{(2)}=0$ only admits the trivial solution.
As a result, all relations in \eqref{eq:SCts_PA} are functionally independent among themselves.

Establishing the functional (in)dependence of
\eqref{eq:SCts_PA} with respect to the primary Lagrangian constraints \eqref{eq:primcon_Pa}
is a more delicate endeavour. 
Consider the row vector (\ref{eq:auxncons}).
In this case, the first nine components are  \eqref{eq:primcon_Pa}, while the latter
three components are \eqref{eq:SCts_PA}.
We search for solutions to \eqref{eq:nullvarphi}. 
A generic ansatz $\bm{\Upsilon}$ has components
given by
\begin{equation}
\label{eq:upsilonPa}
\left(\Upsilon\right)^{ T\leq9}=\Upsilon^{T}_{0}
+\Upsilon^{T}_{1}\,\partial_{1},
\qquad 
\left(\Upsilon\right)^{T>9}=
\Upsilon^{T}_{0},
\qquad
T=1,\dots,12.
\end{equation}
The simplest implementation of \eqref{eq:gammaans} consists in imposing 3 independent conditions, each involving a single $\Upsilon_{0}^{T>9}$ at a time:
\begin{align}
\label{eq:RSec_Pa} 
\sum_{T=1}^{9}
\left(\Upsilon^{T}_{0}
+\Upsilon^{T}_{1}\partial_{1}\right)\phi_T
-\phi_R^{(2)} =0
\qquad
\forall R=1,2,3,
\end{align} 
where, without loss of generality, we have
set $\Upsilon^{T>9}_{0}=-1$.
The advantage of fixing the terms accompanying the secondary constraints is that we can factor the conditions $\boldsymbol{\Upsilon}\cdot \boldsymbol{\Psi}=0$ in a more efficient way.
For instance, evaluating \eqref{eq:RSec_Pa} for $\phi_{1}^{(2)}$ yields the nontrivial solution
\begin{align}
\displaystyle
\bm{\Upsilon}_{1}=
\left( \,0,\, -h,\, -2h^{1},\, 2G_{1}+\partial_{1}, \,G_{11}, \,0, \,2 \mathcal{G}^{1}_{1}, \,\mathcal{G}^{1}_{11}, \, 0,   \, -1, \, 0,\, 0\,\right).
\end{align}
The above implies $\phi^{(2)}_1$
is functionally dependent on the primary Lagrangian constraints.
The analogous strategy for $\phi^{(2)}_{2}$ and $\phi^{(2)}_{3}$ reveals these too are functionally dependent on the primary Lagrangian constraints.
Therefore there are no secondary Lagrangian constraints $l_2=0$ and the algorithm terminates according to closure \ref{item:hardclose}.

For completeness, we present the relations \eqref{eq:SCts_PA} in terms of exclusively the primary Lagrangian constraints:
\begin{align}
\displaystyle
\label{eq:secondaries2D}
\begin{array}{llllll}
&\phi_{1}^{(2)} &\hspace*{-0.2cm}=&\hspace*{-0.2cm}
 \left( 2G_{1}+\partial_{1}\right)\phi_{4}-h{} \,\phi_{2}-2h^1\,\phi_{3}+G_{11} \,\phi_{5}
+2\mathcal{G}^{1}_{1} \,\phi_{7}+\mathcal{G}^{1}_{11}\,\phi_{8}, \vspace*{0.1cm}
\\
&\phi_{2}^{(2)}&\hspace*{-0.2cm}=&\hspace*{-0.2cm}
 \partial_{1} \phi_5+2\left(h\,\phi_{1}-h^{11} \phi_{3} -G \phi_{4}+G_{11} \phi_{6}
-\mathcal{G}^{1} \phi_{7}+\mathcal{G}^{1}_{11} \phi_{9} \right), \vspace*{0.1cm}
\\
&\phi_{3}^{(2)} &\hspace*{-0.2cm}=&\hspace*{-0.2cm}
\left( -2G_{1}+\partial_{1} \right) \phi_6+2h^1 \phi_{1}+h^{11} \phi_{2}-\left(G \phi_{5}
 -\mathcal{G}^{1} \phi_{8}+2\mathcal{G}^{1}_{1} \phi_{9} \right).
\end{array}
\end{align}
Written in this manner,
it becomes obvious that vanishing of the primary Lagrangian constraints implies vanishing of \eqref{eq:SCts_PA} without further restrictions on the field variables $Q^A$.

Using \eqref{eq:masterfor}
and (\ref{eq:lcount}),
we obtain the familiar result
$N_{\textrm{DoF}}=0$.

\end{document}